\pdfoutput=1
\documentclass[12pt]{iopart}

\usepackage[utf8]{inputenc}
\usepackage{lipsum}
\usepackage{graphicx}
\usepackage{pgfplots}
\newcommand{\mom}[1]{\left\langle#1\right\rangle}

\expandafter\let\csname equation*\endcsname\relax
\expandafter\let\csname endequation*\endcsname\relax
\usepackage{amsmath,amssymb}
\newcommand{\sgn}{\operatorname{sgn}}
\usepackage{mathtools}

\usepackage{listings}
\usepackage{xcolor}
\usepackage{tikz,ulem}
\usepackage{bm}
\usepackage{randomwalk}
\usepackage{bm}
\pgfplotsset{width=10cm, height=6cm}
\usetikzlibrary{graphs,graphs.standard}
\usetikzlibrary{decorations.pathreplacing}
\usetikzlibrary{patterns}

\definecolor{dgreen}{rgb}{0,0.7,0}

\def\bluew#1{{\color{blue} #1}}

\usepackage{hyperref}

\newcommand{\Lathrop}[9]{
\begin{scope}
\pgfmathsetmacro{\picwidth}{#1*#2}
\ifthenelse{\equal{#8}{y}}
    {\draw[#9] (0,#6) -- (\picwidth,#6) (0,#7) -- (\picwidth,#7);}
    {}
\draw[#4] (2.55,0.62)
\foreach \x in {1,...,#1}
{   -- ++(#2,rand*#3)
}
coordinate (#5) ;
\end{scope}
\node[right,#4] at (#5) {#5};
}

\usepackage{amsfonts}
\usepackage{graphicx}
\usepackage{hyperref}

\usepackage{xcolor}
\usepackage{subfigure}
\usepackage{textcomp}
\usepackage{cite}
\usepackage{float}
\usepackage{soul}
\usepackage{stackrel}
\usepackage{graphicx}
\graphicspath{{figures_arap/}}

\pdfoutput=1
\usepackage{esint}

\usepackage{color}
\definecolor{dgreen}{rgb}{0,0.7,0}
\def\bluew#1{{\color{red} #1}}

\def\bluew#1{{\color{black} #1}}
\newcommand{\beq}{\begin{equation}}
\newcommand{\eeq}{\end{equation}}
\newcommand{\bea}{\begin{eqnarray}}
\newcommand{\eea}{\end{eqnarray}}

\begin{document}

\title[]{Work statistics at first-passage times}

\author{Iago N Mamede$^1$, Prashant Singh$^2$, Arnab Pal$^3$, Carlos E. Fiore$^1$, and Karel Proesmans$^2$}

\address{$^1$ Universidade de S\~{a}o Paulo,
Instituto de F\'{i}sica,
Rua do Mat\~{a}o, 1371, 05508-090
S\~{a}o Paulo, SP, Brazil}
\address{$^2$ Niels Bohr International Academy, Niels Bohr Institute,
University of Copenhagen, Blegdamsvej 17, 2100 Copenhagen, Denmark}
\address{$^{3}$The Institute of Mathematical Sciences, CIT Campus, Taramani, Chennai 600113, India \& Homi Bhabha National Institute, Training School Complex, Anushakti Nagar, Mumbai 400094, India}
\ead{prashant.singh@nbi.ku.dk, 
arnabpal@imsc.res.in,
karel.proesmans@nbi.ku.dk}

\vspace{10pt}

\begin{abstract}
We investigate the work fluctuations in an overdamped non-equilibrium process that is stopped at a stochastic time. The latter is characterized by a first passage event that marks the completion of the non-equilibrium process. In particular, we consider a particle diffusing in one dimension in the presence of a time-dependent potential $U(x,t)  = k |x-vt|^n/n$, where $k>0$ is the stiffness and $n>0$ is the order of the potential. Moreover, the particle is confined between two absorbing walls, located at $L_{\pm}(t) $, that move with a constant velocity $v$ and are initially located at $L_{\pm}(0) = \pm L$. As soon as the particle reaches any of the boundaries, the process is said to be completed and here, we compute the work done $W$ by the particle in the modulated trap upto this random time. Employing the Feynman-Kac path integral approach, we find that the typical values of the work scale with $L$ with a crucial dependence on the order $n$. While for $n>1$, we show that $\mom{W} \sim L^{1-n}~\exp \left[ \left( {k L^{n}}/{n}-v L \right)/D  \right] $ for large $L$, we get an algebraic scaling of the form $\mom{W} \sim L^n$ for the $n<1$ case. The marginal case of $n=1$ is exactly solvable and our analysis unravels three distinct scaling behaviours: (i) $\mom{W} \sim L$ for $v>k$, (ii) $\mom{W} \sim L^2$ for $v=k$ and (iii) $\mom{W} \sim \exp\left[{-(v-k)L}\right]$ for $v<k$. For all cases, we also obtain the probability distribution associated with the typical values of $W$. Finally, we observe an interesting set of relations between the relative fluctuations of the work done and the first-passage time for different $n$ -- which we argue physically. Our results are well supported by the numerical simulations.
\end{abstract}

\section{Introduction}

Stochastic thermodynamics generally provides a thermodynamic framework to nano-scaled systems subjected to thermal fluctuations, even when they are driven arbitrarily far from the equilibrium \cite{Seifert_2012,VANDENBROECK20156, peliti-stochastic-thermo}. Contrary to classical equilibrium thermodynamics, one now has well-defined probability distributions of the thermodynamic quantities such as heat, work and entropy production \cite{10.1143/PTPS.130.17, PhysRevLett.95.040602}. Furthermore, stochastic thermodynamics can be used to derive several fundamental relations such as fluctuation theorems \cite{PhysRevE.60.2721, PhysRevE.61.2361, PhysRevLett.104.218103, PhysRevLett.78.2690, PhysRevE.50.1645, gallavotti-cohen-1995,gupta2020work}, uncertainty relations \cite{PhysRevLett.114.158101, PhysRevLett.116.120601, Proesmans_2017, PhysRevLett.123.110602, PhysRevLett.123.090604, PhysRevLett.122.230601, ProesmansTUR_2019, HarunariTUR_2020, PhysRevResearch.2.022044, PhysRevResearch.3.013273,PhysRevResearch.3.L032034} and thermodynamic speed limits \cite{PhysRevLett.125.120604, dLimmer, PhysRevLett.128.050603, PhysRevLett.121.070601, PhysRevLett.106.250601, PhysRevX.10.021056, aurell-et-al-2012, PhysRevLett.127.190602, PhysRevLett.108.190602, PhysRevLett.125.100602, PhysRevE.102.032105, PhysRevLett.128.010602, Dechant_2022}.

Recently, there has been a growing interest in the study of the thermodynamic quantities until a particular event of interest has occurred for the first time \cite{PhysRevX.7.011019, Chetrite_2018, PhysRevLett.122.220602, Neri_2019, PhysRevLett.124.040601, chetrite-gupta-2011, 10.21468/SciPostPhys.12.4.139, PhysRevLett.115.250602, Nerii_2022, PhysRevLett.122.230602, PhysRevE.105.024112, 10.21468/SciPostPhys.14.5.131, PhysRevResearch.3.L032034}. Examples include the escape of a colloidal particle from a metastable state or stretching of a polymer till a certain length is attained \cite{PhysRevLett.124.040601}. Since the underlying dynamics is stochastic, the time at which these events take place also varies from realisation to realisation. This drastically changes the properties of the thermodynamic quantities compared to situations where the observation time is fixed for all realisations. For instance, the bound on the average work picks up a non-trivial correction term due to the fact that the system is generally out of equilibrium at the end of the first-passage time \cite{PhysRevLett.124.040601}. Similar observation has also been made for the efficiency of heat engines \cite{Neri_2019}. In fact, based on the martingale theory, many new results on the integral fluctuation theorem and  stopping times for entropy production have been analytically obtained \cite{roldan-et-al-2023}. 
Despite these general results, the number of thermodynamic first-passage problems that have been solved exactly, seems to be very limited.
In this paper, we aim to partially fill this gap by studying an analytically tractable class of models, where one can get exact results for the moment generating function associated with the mechanical work.

In order to study this, we reformulate the work done as a first-passage functional of the stochastic process \cite{Majumdar-review-BF}. Statistical properties of such functionals have been quite extensively studied in the literature by deploying the celebrated Feynman-Kac formalism suitably adapted for first-passage problems \cite{92dffd60-7698-3838-9cfd-f3a5e322ac6b}. Based on this formalism, these functionals have been shown to have many applications in fields ranging from queue theory \cite{MJKearney_2004}, sandpile and percolation models \cite{PhysRevLett.63.1659, prellberg1995critical} to disordered systems \cite{DavidSDean_2001}, among others \cite{richard2002scaling, PhysRevE.76.031130, Bray2013}. Moreover, they have been studied for diverse stochastic processes such as diffusion and drift-diffusion \cite{abundo2013firstpassage, Kearney_2005, Majumdar_2020, abundo2017jointdistribution}, random acceleration \cite{PhysRevE.107.064122}, L\'{e}vy process \cite{abundo2019jointdistribution}, Ornstein-Uhlenbeck particle \cite{Kearney_2021, mario2022bm} and resetting processes \cite{Singh_2022_FP, Dubey_2023}. In this paper, we are interested in using these tools and techniques to calculate the moments and the distribution of the work done by a diffusing particle subjected to a time-dependent potential $U(x,t)  = \frac{k}{n} |x-vt|^n$ with order $n >0$. In particular, we show that the work can be reformulated as a first-passage functional and one can then employ the Feynman-Kac formalism to investigate its statistical properties.

The rest of the paper is structured as follows: In Section \ref{model}, we introduce our model and present a general formalism to study the work done. We first focus on the analytically tractable cases of the linear potential $(n=1)$ in Section \ref{linear} and the harmonic potential ($n=2$) in Section \ref{harmonic}. The insights gained from these two cases become instrumental in dealing with the general $n$ in Section \ref{general}. We then discuss an interesting relation between the work done and the first-passage time in Section \ref{firstpassagetime}. Finally, in Section \ref{conclusion}, we present the conclusion and make some future remarks.

\begin{figure}[t]
    \centering
    \includegraphics[scale=0.3]{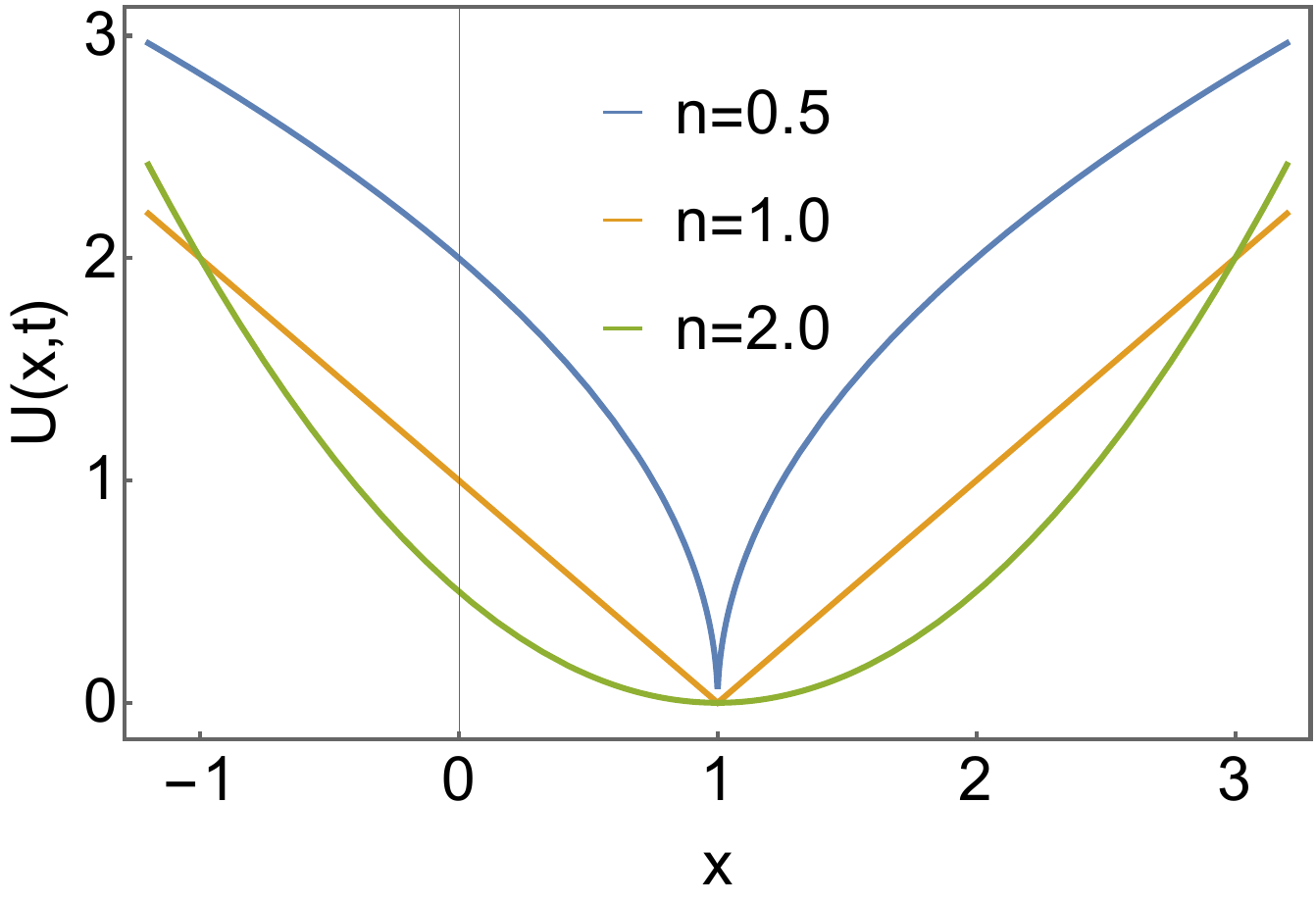}
    \includegraphics[scale=0.3]{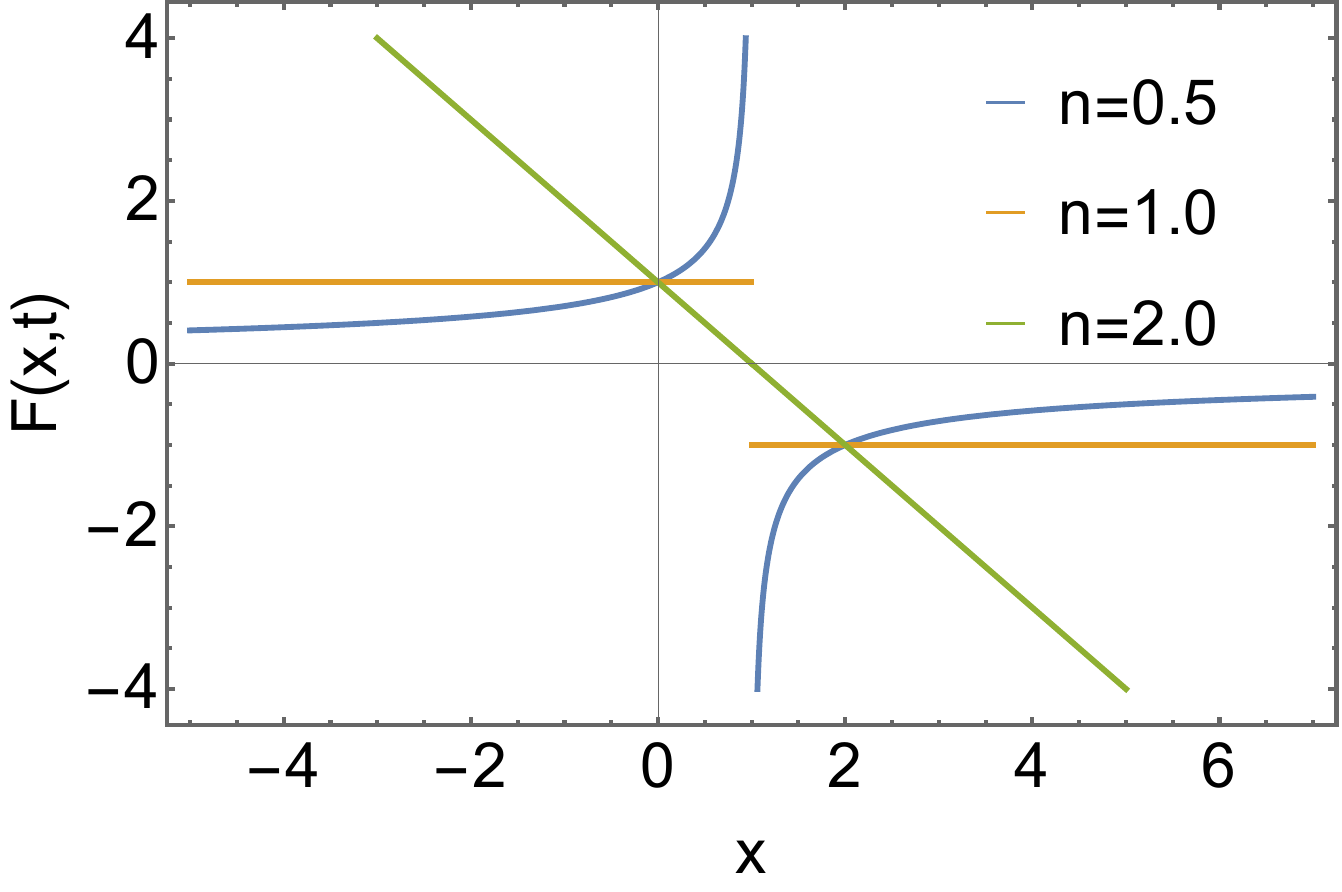}
    \caption{Left and right panels: The depiction of the potential $U(x,t)$ (via Eq.~\eqref{pott-eqn}) and the corresponding force $F(x,t)$ for different values of $n$. Parameters: $k=v=t=1$.}
    \label{sketch-fig}
\end{figure}
\section{Model and general formulation of the problem}\label{model}
We consider a one-dimensional diffusing particle whose position, in presence of a time-dependent potential $U(x,t)$, evolves according to an overdamped Langevin equation 
\bluew{\begin{align}
\frac{dx}{dt} = - \frac{1}{\bar{\gamma}}\frac{\partial U(x,t)}{ \partial x} + \sqrt{2 D}~\eta(t),
\label{langevin}
\end{align}
where $\eta(t)$ is the Gaussian white noise satisfying the properties $\langle \eta(t)\rangle=0$ and $\langle \eta(t)\eta(t')\rangle=\delta(t-t')$. Also $D$ and  ${\bar \gamma}$ denote the diffusion and damping coefficients, respectively, and are related  by the Einstein relation $D=k_BT/{\bar \gamma}$. Throughout this paper, we will fix $\bar{\gamma}=1$.} The initial position of the particle will be denoted as $x_0$. Moreover, we choose the potential $U(x,t)$ to be a confining potential with stiffness $k~(>0)$ and we let it move with a constant velocity $v~(>0)$:
\begin{align}
U(x,t) = \frac{k}{n} ~\big|x-v t\big| ^{n},~~~~~\text{with }n >0. \label{pott-eqn}
\end{align}
Finally, we will assume the existence of two absorbing walls located at positions $L_{\pm}(t)$ that themselves move with the same constant velocity $v$, i.e. $L_{\pm}(t) = \pm L +  v t$, with $-L<x_0<L$. \bluew{Plots of the potential $U(x,t)$ and the associated force $F(x,t)$ are provided in Figure \ref{sketch-fig}. We observe that for $n < 1$, the magnitude of the force decays with increasing $x$, while for $n > 1$, it increases with $x$. 
Intriguingly, we observe different striking behaviour depending on the order of the potential strength as will be illustrated in this paper.}


As mentioned in the introduction, we are interested in the work distribution of first-passage time problems. Here, we allow the particle to move until it touches one of the absorbing boundaries at $L_{\pm}(t)$. Let us denote this time by $t_f$ and the corresponding trajectory as $\{ x(\tau) \}$ with $0 \leq \tau \leq t_f$. We then calculate the work done up to time $t_f$ as \cite{10.1143/PTPS.130.17, PhysRevE.67.046102, PhysRevLett.91.110601, jarzynski1999} 
\begin{align}
        W(x_0)& = \int_{0}^{t_f}d\tau~\frac{\partial U(x(\tau), \tau)}{\partial \tau}= - k v  \int_{0}^{t_f}d\tau~\sgn\Big(x(\tau)-v \tau \Big)~\big|x(\tau)-v \tau\big|^{n-1}, \label{work} \\
       \text{with          }t_f&  =\text{min}\{\ \tau:x(\tau) = L_{+}(\tau),~\text{or } x(\tau) = L_{-}(\tau)\}. \label{FPT-new}
\end{align}
Since the motion is random, we get different values of $t_f$ for different realisations. This indicates that in contrast to the usual case of fixed observation time, the definition of work in Eq.~\eqref{work} possesses two sources of stochasticity - one arises due to the random trajectory $\{ x(\tau) \}$ and the other stems from random first-passage time $t_f$.

\bluew{Notice that we have considered the absorbing boundaries to be dynamical in our model. Such scenarios might arise in experiments in which a Brownian particle is dragged using a moving potential \cite{PhysRevE.67.046102, PhysRevLett.91.110601}. Typically, they have been carried out for a fixed observation time. However, one can also consider scenarios where one waits till the particle comes out of the field of view of the camera or escapes from the trap. Now if the camera/trap is moving, the field of view also changes with time and these, in turn, mimic a moving boundary scenario. Another example might be investigating the statistics of work for stretching a polymer to a certain threshold length that itself depends on time  \cite{PhysRevLett.124.040601}. This can also be formulated as a moving boundary problem.}


In order to derive the statistics of $W(x_0)$, we reformulate the work done as a first-passage functional which has been extensively studied in the literature \cite{Majumdar-review-BF}. To see this, let us first perform a change of variable as
\begin{align}
\xi (t) = x(t)- v t,
\end{align}
and recast the Langevin equation and the work done in terms of this new co-moving variable:
\begin{align}
&  \frac{ \partial \xi}{\partial t} =  -k \sgn\left( \xi(t) \right)\big|\xi(t)\big|^{n-1}- v +\sqrt{2D}~\eta(t), \label{LangevinNew} \\
&   W(x_0) = - k v  \int_{0}^{t_f}d\tau~\sgn\Big( \xi(\tau) \Big)~|\xi(\tau)|^{n-1}. \label{ajha}
\end{align}
Meanwhile, the first-passage time $t_f$, in the co-moving frame, simply becomes the time at which the particle moves out of the fixed interval $[-L, L]$ for the first time. Therefore, by suitably defining the variable $\xi(t)$, we have been able to recast the entire problem with fixed absorbing walls. Following \cite{92dffd60-7698-3838-9cfd-f3a5e322ac6b, Majumdar-review-BF}, one can now derive a backward differential equation for the moment generating function
\begin{align}
    \centering
    Q(p,\xi_0)= \mom{e^{-p W(\xi _0)}} = 
    \mom{\exp\left[k p v\int_{0}^{t_f} d\tau~\sgn\Big( \xi(\tau) \Big)~\big| \xi(\tau)\big| ^{n-1}\right]}, \label{mom-gen}
\end{align}
with $\xi _0 = \xi(t=0)=x_0$ and the averaging $\mom{...}$ involves averaging both with respect to the trajectories as well as the first-passage time $t_f$. To do this, we now look at a trajectory of $\xi (\tau)$ from $\tau = 0$ to $\tau = t_f$ and break it into two parts: (i) a left interval $[0, \Delta t]$ and (ii) a right interval $[\Delta t, t_f]$ with $\Delta t \to 0$. At the end of the left interval, the variable $\xi (t)$ takes the value $\xi(\Delta t) = \xi _0 -\left[ k \sgn(\xi _0) |\xi _0 |^{n-1}+v  \right] \Delta t + \sqrt{2 D} ~\Delta t ~\eta(0)$. In the remaining time interval $(t_f - \Delta t)$, the particle starting from $\xi(\Delta t) $ moves out of the interval $[-L, L]$. Therefore, decomposing the integral in Eq.~\eqref{mom-gen} as $\int _0^{t_f} = \int _{0}^{\Delta t} + \int _{\Delta t}^{t_f}$, we obtain
\begin{align}
Q(p,\xi_0) &  \simeq \mom{ \exp \left[kpv \sgn(\xi _0)  \big|\xi _0 \big|^{n-1}  \Delta t \right] ~Q(p, \xi(\Delta t)) }, \\
& \simeq  \left[1+kpv \sgn(\xi _0)  \big|\xi _0 \big|^{n-1}  \Delta t \right]~\mom{Q(p, \xi(\Delta t)) }.
\end{align}


Inserting the expression of $ \xi(\Delta t)$ and performing the expansion in $\Delta t$, we obtain the following differential equation for $Q(p,\xi_0) $ as $\Delta t \to 0$:
\begin{align}
D \frac{\partial ^2 Q(p,\xi_0)}{\partial \xi _0 ^2} - \Big(k f_n(\xi _0)+v \Big) \frac{\partial Q(p,\xi_0) }{\partial \xi_0} + kpv f_n(\xi _0) ~Q(p,\xi_0)  = 0, \label{BackWard}
\end{align}
where $f_n(\xi _0) =\sgn(\xi _0) \big|\xi _0 \big|^{n-1} $. This equation has to be solved in the domain $-L \leq \xi _0 \leq L$ and it a differential equation in the initial position $\xi _0$. This approach is referred to as a``backward" approach as it involves varying the initial position instead of the final one. Moreover it is a second order differential equation. Thus, we need two boundary conditions to solve it. These conditions can be obtained by analysing the behaviour of $Q(p,\xi_0)$ at $\xi _0 = \pm L$. For these two cases, the particle is already at the edge of the interval and gets instantly absorbed. Thus, for both cases, we have $t_f = 0$ which implies $W(\pm L) = 0$ from Eq.~\eqref{ajha}. Plugging this in Eq.~\eqref{mom-gen}, we obtain
\begin{align}
Q(p,\xi_0 = \pm L) = 1. \label{Bcs}
\end{align}
The goal now is to solve the backward Eq.~\eqref{BackWard} with these boundary conditions and then utilize it to obtain the moments of $W(x_0)$ as
\begin{align}
    \mom{W^m(\xi _0)} = (-1)^m \frac{\partial^m Q(p, \xi_0)}{\partial p^m}\bigg|_{p=0}.
    \label{Moments}
\end{align} 

One can also obtain the full probability distribution by performing the inverse transformation with respect to the $p$-variable in Eq.~\eqref{mom-gen}. In what follows, we first illustrate this rigorously for the analytically tractable cases $n=1$ and $n=2$. Then, we will combine the intuition gained in these solvable cases with Eq.~\eqref{BackWard} to derive some general results for arbitrary $n$.

\section{Linear potential $(n=1)$}\label{linear}
\quad  Let us first look at the $n = 1$ case for which we can solve the backward Eq.~\eqref{BackWard} analytically. 
For this case, the Langevin equation \eqref{LangevinNew} takes the form
\begin{align}
&  \frac{ \partial \xi}{\partial t} =  -k \sgn\left( \xi(t) \right)- v +\sqrt{2D}~\eta(t).
\end{align}
In essence, this is a drift-diffusion process in which the particle experiences two distinct drifts. The first one involves a drift with a magnitude $k$, that tries to pull  the particle back to the origin. The second one involves a drift of magnitude $v$ that tends to take the particle away from the origin along the negative $x$ direction. The behaviour of the particle varies depending on which of the two terms dominate. To see this in the context of work done, let us proceed to solve the backward Eq.~\eqref{BackWard}. Depending on the sign of the initial position $\xi _0$, we can split the backward equation as
\begin{align}
    \begin{cases}
        \left[ \partial^2_{\xi_0}-\mu\partial_{\xi_0}+\frac{k v p}{D} \right]Q(p,\xi_0)=0\text{ ,    for }\xi_0 > 0\\
        \left[ \partial^2_{\xi_0}-\gamma\partial_{\xi_0}-\frac{k v p}{D}\right] Q(p,\xi_0)=0\text{ ,    for }\xi_0 < 0,
    \end{cases}
\end{align}
\noindent where $\mu = (v+k)/D$, $\gamma = (v-k)/D$. Solving this set of equations yields
\begin{align}
    \centering
    Q(p,\xi_0)&=\begin{cases}
        K_1(p)e^{\lambda_+(p)\xi_0}+K_2 (p)e^{\lambda_-(p)\xi_0}\text{ , for }\xi_0 > 0\\
         K_3(p)e^{\sigma_+(p)\xi_0}+K_4(p)e^{\sigma_-(p)\xi_0}\text{ , for }\xi_0 < 0,
        \label{QLinear}
    \end{cases}\\
     \text{with   } \lambda _{\pm}(p)&  = \frac{1}{2} \left( \mu \pm \sqrt{\mu ^2 -\frac{4k v p}{D}} \right)~\text{and  }\sigma _{\pm}(p)  = \frac{1}{2} \left( \gamma \pm \sqrt{\gamma ^2 +\frac{4k v p}{D}} \right). \label{lamb-eq}
\end{align}
To compute the $K(p)$-functions, we will need four conditions. Two of these are the boundary conditions $Q(p, \xi _0 = \pm L)$ which are given in Eq.~\eqref{Bcs}. The other two can be obtained by looking at the behaviour of $Q(p, \xi _0)$ and its derivative in the vicinity of $\xi _0 = 0$. Integrating the backward Eq.~\eqref{BackWard} from $-\epsilon $ to $+ \epsilon$ and then taking $\epsilon  \to 0^+$, we obtain
\begin{align}
    \centering
    \begin{cases}
        Q(p, \xi _0 \to 0^+)=Q(p, \xi _0 \to 0^-),\\
        \partial_{\xi_0}Q(p,\xi _0) \big|_{\xi _0 \to 0^+}=\partial_{\xi_0}Q(p, \xi _0) \big|_{\xi _0 \to 0^-}.
    \end{cases}
\end{align}
Using these conditions, it is possible to compute all $K(p)$-functions as
\footnotesize{\begin{align}
    K_1(p)&=\frac{\left(\sigma _{-}-\sigma _{+}\right) e^{\lambda _ - L} +\left(\sigma _{+}-\lambda _{-}\right) e^{-\sigma _-L}+\left(\lambda _{-}-\sigma _{-}\right)e^{-\sigma _+ L}}{e^{\lambda _+ L}[\left(\sigma _{+}-\lambda _{-}\right)  e^{-\sigma _- L}+\left(\lambda _{-}-\sigma _{-}\right)  e^{-\sigma _+L}]+e^{\lambda _ -L}[\left(\lambda _{+}-\sigma _{+}\right)  e^{-\sigma _- L}+\left(\sigma _{-}-\lambda _{+}\right)  e^{-\sigma _+ L}]},    \nonumber \\
   K_2(p)&=\frac{\left(\sigma _{+}-\sigma _{-}\right) e^{\lambda _+L} +\left(\lambda _{+}-\sigma _{+}\right) e^{-\sigma _- L}+\left(\sigma _{-}-\lambda _{+}\right)
  e^{-\sigma _+ L}}{e^{\lambda _+ L}[\left(\sigma _{+}-\lambda _{-}\right)  e^{-\sigma _- L}+\left(\lambda _{-}-\sigma _{-}\right)  e^{-\sigma _+L}]+e^{\lambda _ -L}[\left(\lambda _{+}-\sigma _{+}\right)  e^{-\sigma _- L}+\left(\sigma _{-}-\lambda _{+}\right)  e^{-\sigma _+ L}]},   \nonumber \\
  K_3(p) & =  \frac{     \left(\lambda _{+}-\sigma _{-}\right) e^{\lambda _-L} +          \left(\sigma _{-}-\lambda _{-}\right) e^{\lambda _+ L}+         \left(\lambda _{-}-\lambda _{+}\right)
  e^{-\sigma _- L}          }{            e^{\lambda _+ L}[\left(\sigma _{+}-\lambda _{-}\right)  e^{-\sigma _- L}+\left(\lambda _{-}-\sigma _{-}\right)  e^{-\sigma _+L}]+e^{\lambda _ -L}[\left(\lambda _{+}-\sigma _{+}\right)  e^{-\sigma _- L}+\left(\sigma _{-}-\lambda _{+}\right)  e^{-\sigma _+ L}]}, \nonumber \\
   K_4(p) & =  \frac{     \left(\lambda _{+}-\sigma _{+}\right) e^{\lambda _-L} +          \left(\sigma _{+}-\lambda _{-}\right) e^{\lambda _+ L}+         \left(\lambda _{-}-\lambda _{+}\right)
  e^{-\sigma _+ L}          }{            e^{\lambda _+ L}[\left(\sigma _{+}-\lambda _{-}\right)  e^{-\sigma _- L}+\left(\lambda _{-}-\sigma _{-}\right)  e^{-\sigma _+L}]+e^{\lambda _ -L}[\left(\lambda _{+}-\sigma _{+}\right)  e^{-\sigma _- L}+\left(\sigma _{-}-\lambda _{+}\right)  e^{-\sigma _+ L}]}, \nonumber
\end{align}}
\normalsize{and} inserting them in Eq.~\eqref{QLinear}, we get for $\xi _0 = 0$
\begin{align}
Q(p,0) = K_1(p)+K_2(p). \label{mom-eq-1}
\end{align}
This gives us the exact form of the moment-generating function for the case of linear potential. By taking derivative of Eq.~\eqref{mom-eq-1} suitably with respect to $p$, we can now obtain all moments of the work $W$. For instance, we find the mean as
\begin{figure}[t]
    \centering
    \includegraphics[scale=0.22]{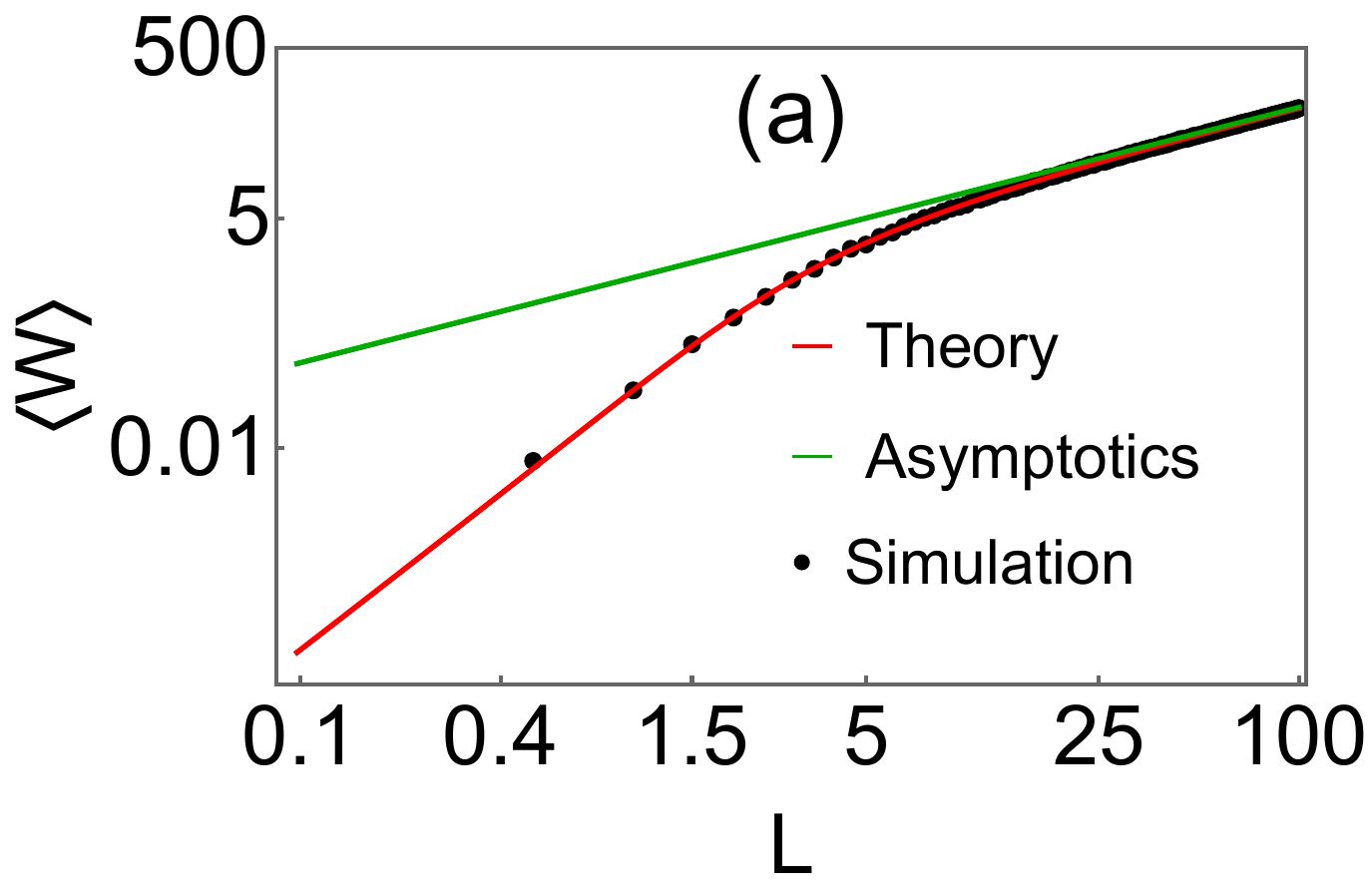}
    \includegraphics[scale=0.22]{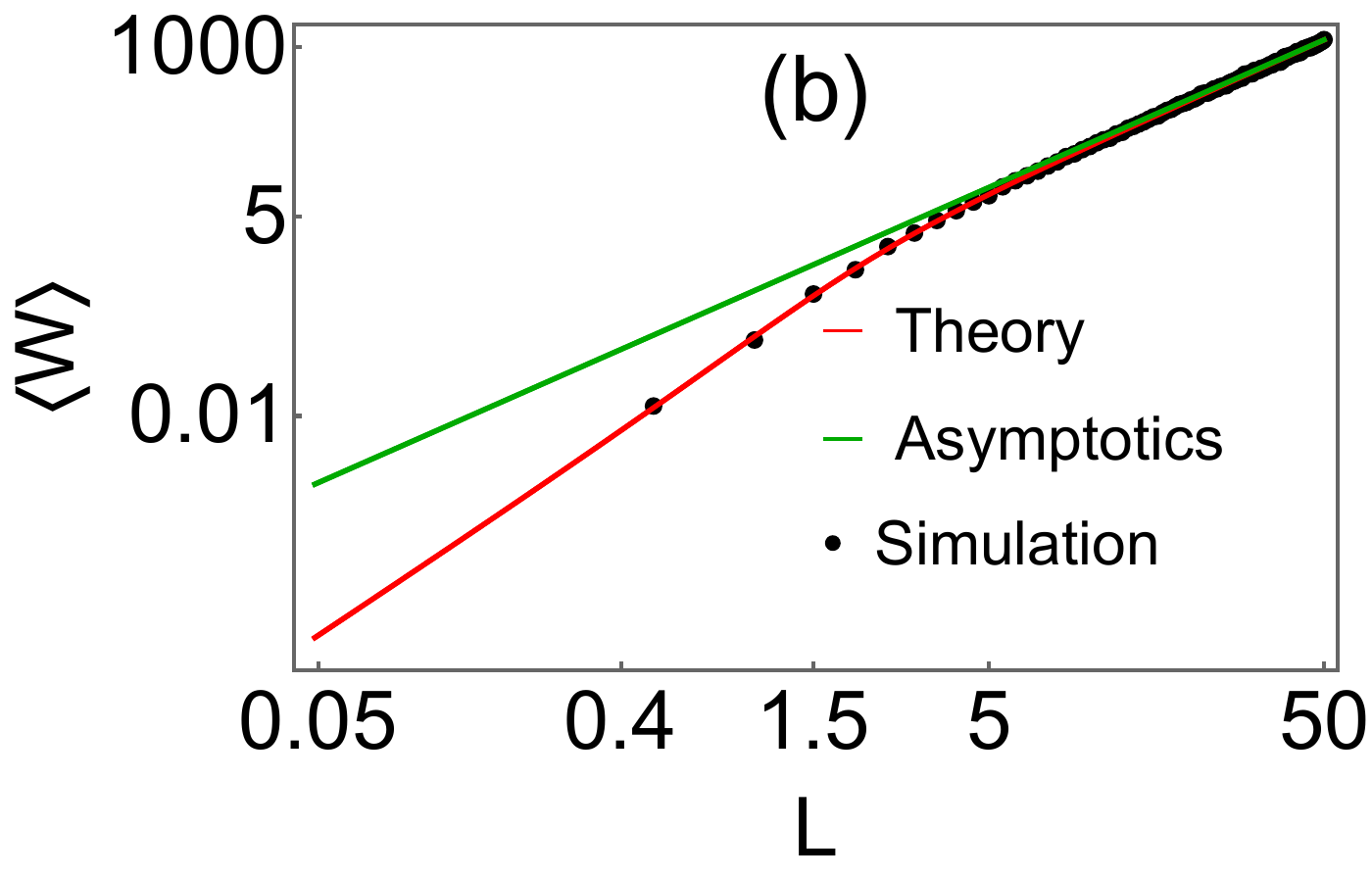}
    \includegraphics[scale=0.21]{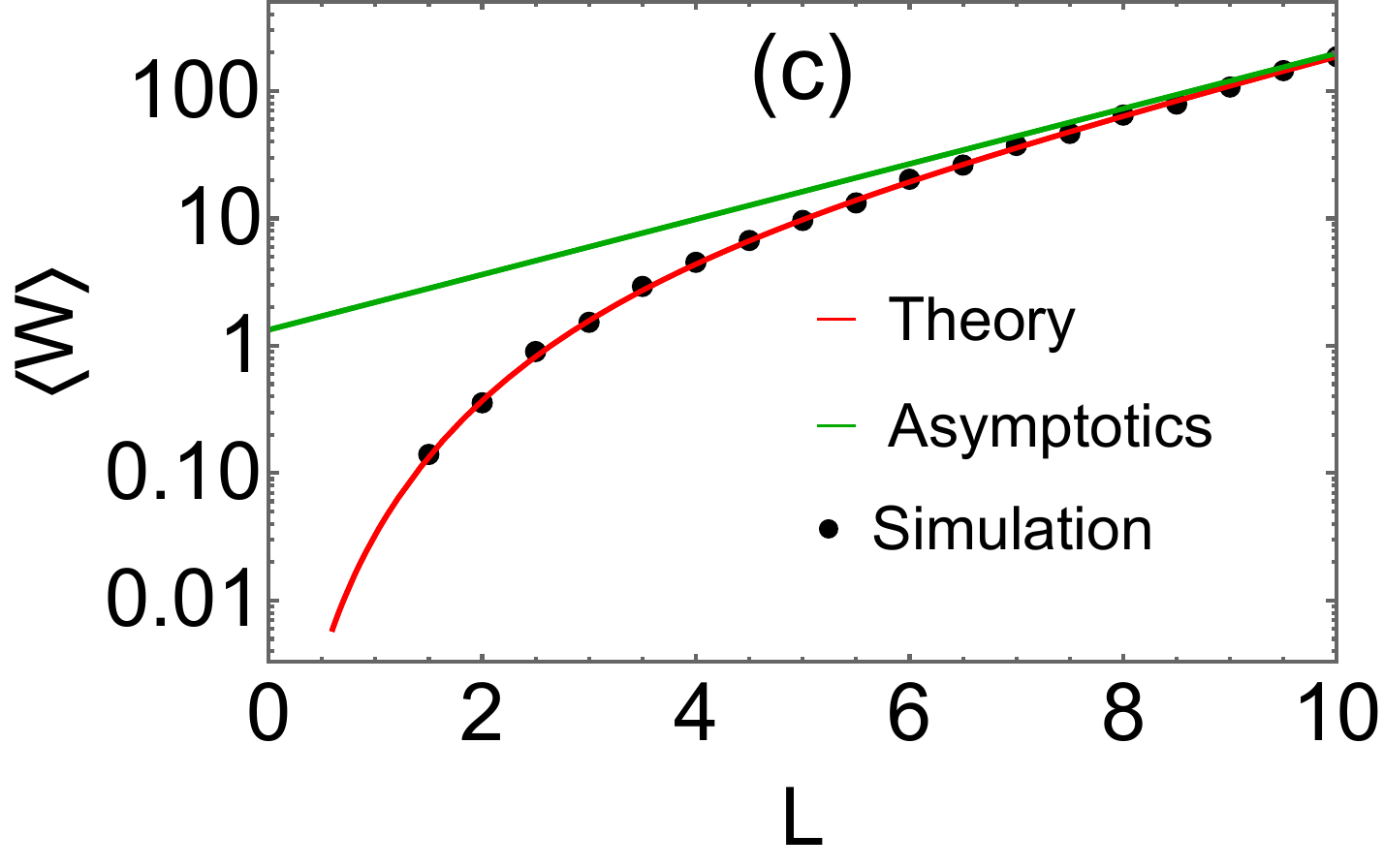}
    \caption{Plots of the mean work $\mom{W}$ for the linear $n=1$ potential for (a) $\gamma >0~(v=1,~k=0.5)$, (b) $\gamma =0~(v=k=1)$ and $\gamma <0~(v = 0.5,~k=1)$. For all panels, the red curve corresponds to the exact result in Eq.~\eqref{mom-eq-2} and the green curve is the asymptotic expression in Eq.~\eqref{mom-eq-3}. We have chosen $D=1$.}
    \label{fig-ps-1}
\end{figure}
\begin{align}
    \centering
    \mom{W}=\left(\frac{kv}{\gamma {D} \mu}\right) \frac{(\gamma + \mu) \left[ e^{\mu L}+e^{-\gamma L}-e^{(\mu-\gamma)L}-1  \right] - \gamma \mu L \left[ e^{\mu L}-e^{-\gamma L}\right]}{\gamma \left[1- e^{\mu L}\right] - \mu \left[ 1- e^{-\gamma L} \right]}. \label{mom-eq-2}
\end{align}
Here, we have introduced the simplified notation $\mom{W^m}$ instead of $\mom{W^m(0)}$. To gain some physical insights, let us look at the asymptotic behaviour of this mean as $L$ becomes large. This behaviour turns out to crucially depend on the signature of the parameter $\gamma$ and we find
\begin{align}
\mom{W} \simeq 
\begin{cases}
\frac{k v L}{\gamma D},~~~~~~~~~~~~~~\text{for }\gamma >0, \\
\frac{v^2 L^2}{2 D},~~~~~~~~~~~~~\text{for }\gamma =0, \\
\frac{2 k v^2}{\gamma ^2 D^2 \mu}~e^{-\gamma L},~~~~\text{for  }\gamma <0.
\end{cases}
\label{mom-eq-3}
\end{align}
Physically, for $\gamma <0$ (or equivalently $v < k$), the drift $v$ in Eq.~\eqref{LangevinNew} is not strong enough for the particle to overcome the force $k$. Thus, the particle typically takes exponentially large time to reach the absorbing walls which, in turn, gives rise to exponentially large value of the work. On the other hand, for $\gamma >0$ (or equivalently $v > k$), the particle can easily overcome the constant force and reach the absorbing walls with relatively high probability. Therefore, we get smaller values of the work which scales linearly with $L$. For the marginal case $\gamma =0$, we anticipate the scaling to lie somewhere in between $\gamma >0$ and $\gamma <0$ cases. In Figure~\ref{fig-ps-1}, we have  compared these asymptotic results and their exact expressions in Eq.~\eqref{mom-eq-2} with the numerical simulation of the Langevin equation~\eqref{langevin}.

After analysing the mean, we now proceed to compute the full probability distribution of the work. For this, we need to perform the inverse Laplace transformation of $Q(p,0)$ in Eq.~\eqref{mom-eq-1}. For large $L$, we saw that the work typically takes a large positive value for all values of $\gamma$. In terms of $p$-variable, this translates to taking small $p$ limit of the moment generating function. Taking $p \to 0$ limit in Eq.~\eqref{mom-eq-1}, we find that $Q(p,0)$ also takes different forms depending on the signature of $\gamma$. In \ref{appen-linear}, we have rigorously derived these forms for large $L$ and found
\begin{align}
Q(p,0) \simeq 
\begin{cases}
\exp \left(  \frac{\gamma L}{2} -\frac{L}{2}\sqrt{\gamma ^2 + \frac{4vk p}{D}}\right),~~~~~~~~~~~~~~~~~~~~\text{for }\gamma >0, \\
\\
\frac{\mu \left\{ \exp(2 \alpha L \sqrt{p}) -1  \right\} +2 \beta \sqrt{p} \exp(\alpha L \sqrt{p}) }{\mu \left\{ \exp(2 \alpha L \sqrt{p}) -1  \right\} + \beta \sqrt{p} ~ \left\{ \exp(2 \alpha L \sqrt{p}) +1  \right\} },~~~~~~~~~~\text{for }\gamma =0, \\
\\
\left[ 1+p \mom{W} \left( 1+ \frac{k v L}{D \big| \gamma \big|}p\right)  \right]^{-1},~~~~~~~~~~~~~~~~~\text{for  }\gamma <0,
\end{cases}
\label{mom-eq-4}
\end{align}
where $\alpha = v /\sqrt{D}$ and $\beta = \alpha \left( e^{\mu L}-1 \right)$. For $\gamma \neq 0$, we can perform the inverse Laplace transformation explicitly and obtain the distribution of work to be
\begin{align}
P(W) \simeq
\begin{cases}
L \sqrt{\frac{k v}{4 \pi D W^3}} \exp \left[ -\frac{D\gamma ^2}{4 k v W} \left( W - \frac{k v L}{D} \right)^2   \right],~~~~~\text{for } \gamma >0, \\
\\
\frac{1}{\mom{W}} \left[  \exp \left( -\frac{W}{\mom{W}}\right)  -\exp \left( -\frac{D \big| \gamma \big| W}{k v L}  \right) \right],~~~~\text{for } \gamma <0.
\end{cases}
\label{mom-eq-5}
\end{align}
\begin{figure}[t]
    \centering
    \includegraphics[scale=0.22]{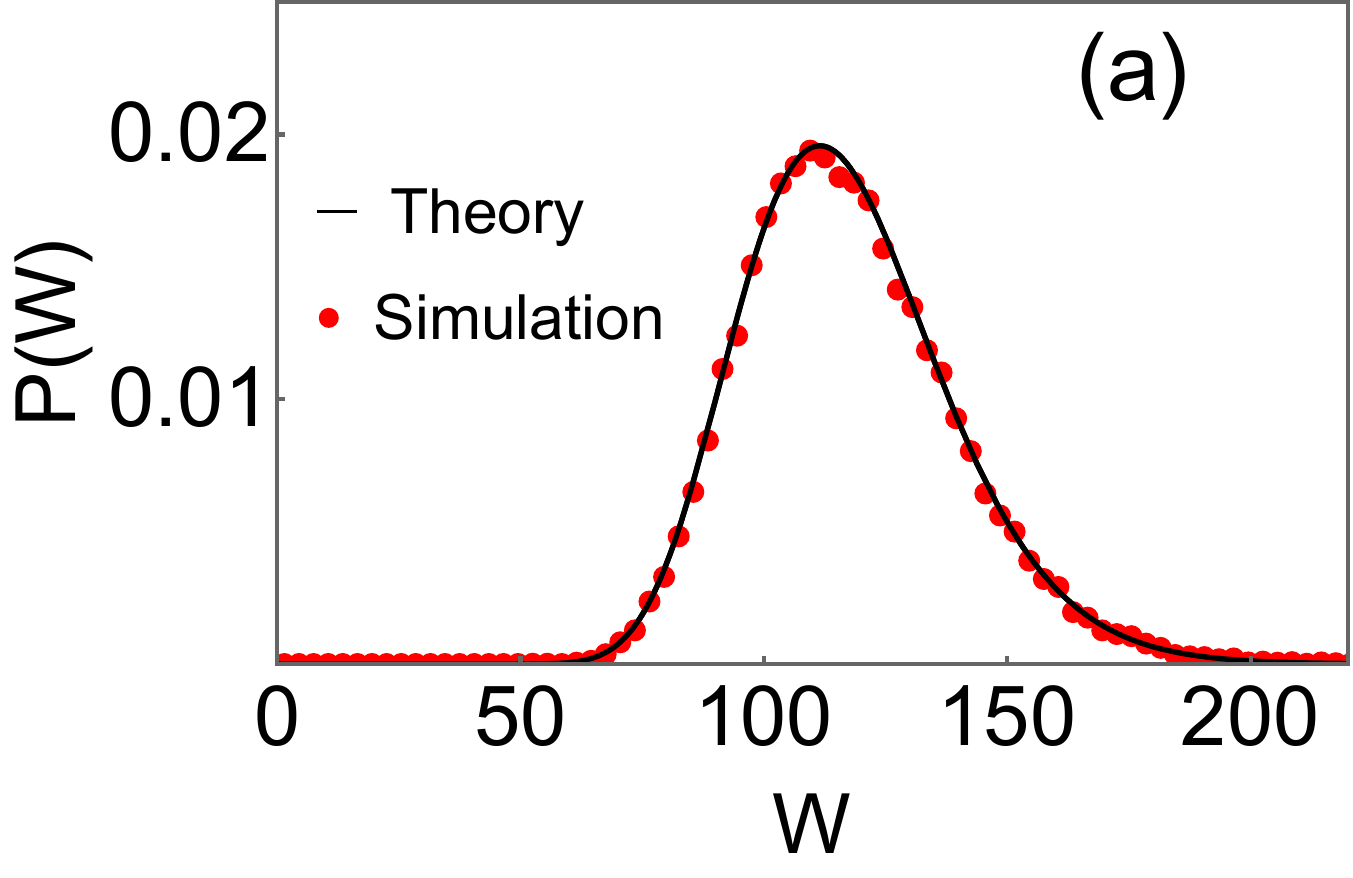}
    \includegraphics[scale=0.215]{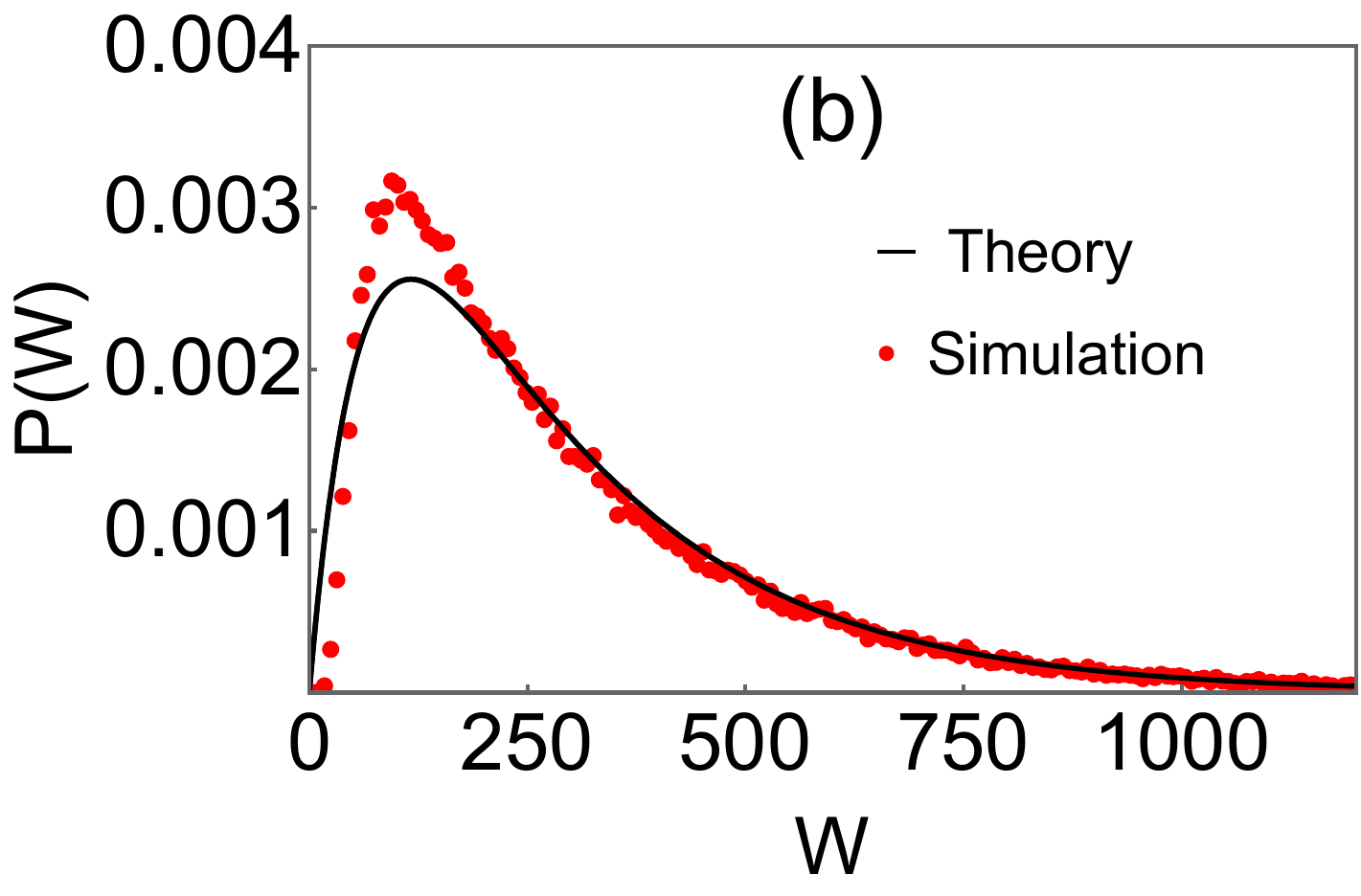}
    \includegraphics[scale=0.225]{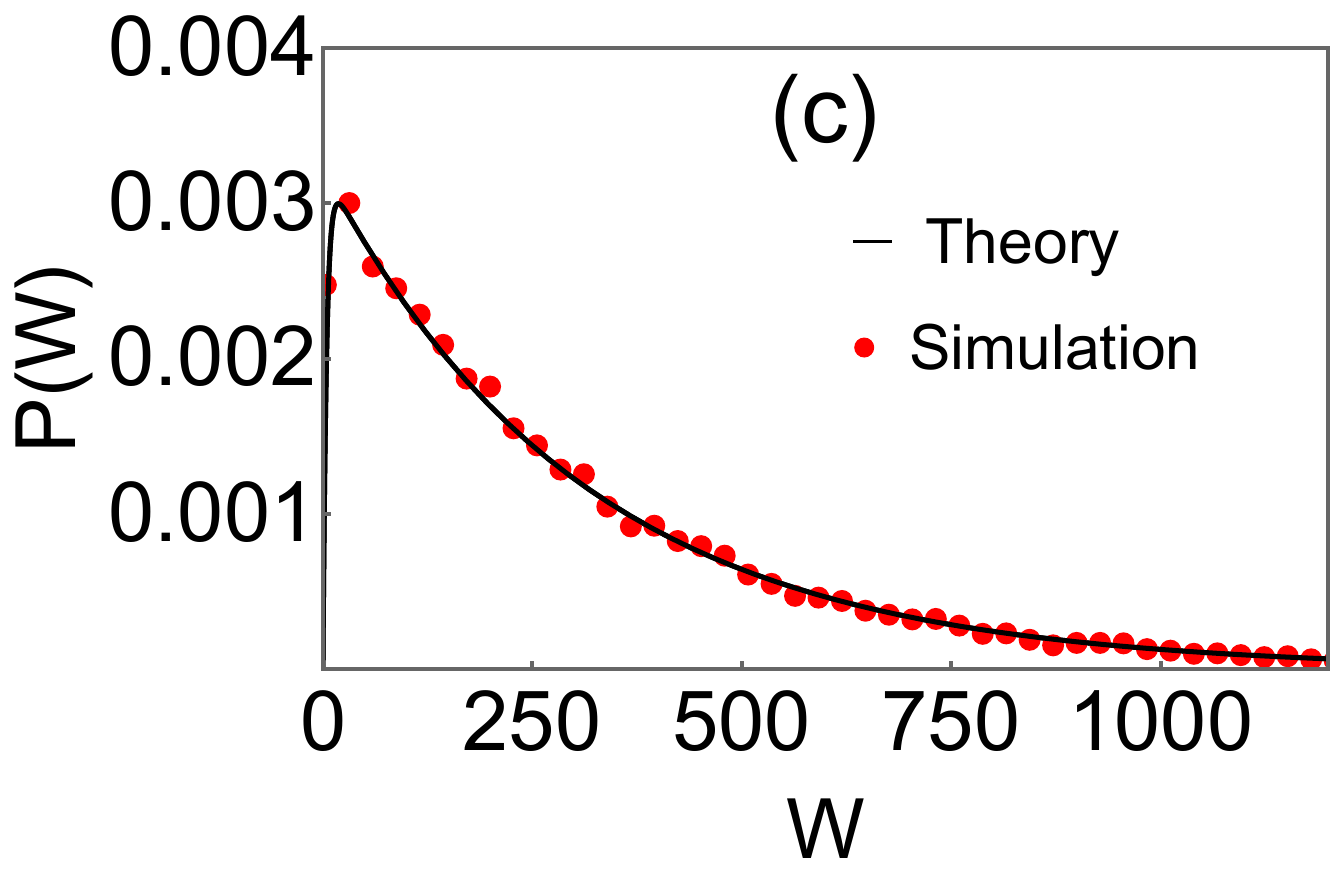}
    \caption{Distribution of the work done by the particle for the linear potential $(n=1)$ for (a) $\gamma >0~(v=1,~k=0.5,~L=120)$, (b) $\gamma =0~(v=k=1,~L=25)$ and $\gamma <0~(v = 0.3,~k=1,~L=10)$. The theoretical expressions are given in Eqs.~\eqref{mom-eq-5} and \eqref{mom-eq-7}. We have chosen $D=1$.}
    \label{fig-ps-2}
\end{figure}
On the other hand, for $\gamma =0$, we perform the expansion
\begin{align}
Q(p, 0 ) \simeq \frac{1}{1+\frac{L^2 \alpha ^2}{2}p + \frac{L^4 \alpha^4}{24} p^2}, \label{mom-eq-6}
\end{align}
and then carry out the inversion with respect to $p$ to obtain
\begin{align}
P(W) \simeq \frac{4 \sqrt{3}}{L^2 \alpha ^2}~\exp \left( -\frac{6W}{L^2 \alpha ^2}  \right)~\sinh \left( \frac{2 \sqrt{3} W}{L^2 \alpha ^2} \right), ~~~~\text{for }\gamma =0. \label{mom-eq-7}
\end{align}
In Figure~\ref{fig-ps-2}, we have compared our analytic results with the same obtained from numerical simulations. We observe excellent match between the derived results and numerics for $\gamma \neq 0$. Contrarily for $\gamma =0$, we see a departure at smaller values of $W$ [see middle panel in Figure~\ref{fig-ps-2}]. This stems from the fact that we have truncated the series in Eq.~\eqref{mom-eq-6} at $p^2$ and ignored the higher order terms. While this is valid for large values of $W$ (which in the Laplace domain corresponds to small $p$), it becomes less accurate for smaller values of $W$ (which corresponds to large $p$). Thus, for a more accurate match at small $W$, one needs to consider higher order terms in Eq.~\eqref{mom-eq-6}. 

Furthermore, the distribution of $W$ for these different cases turn out to be completely different. For instance, as seen in panel (a) in Figure~\ref{fig-ps-2} for $\gamma >0$, the distribution close to the mean can be effectively approximated by a Gaussian form. This can also be seen in Eq.~\eqref{mom-eq-5} where we can replace $W \simeq \mom{W}$ in the vicinity of the mean and the distribution then simply becomes a Gaussian distribution. However, as we move towards the tail, this approximation ceases to remain valid and one needs to consider the full non-Gaussian form of the distribution. Contrarily, for $\gamma \leq 0$, the distribution is strictly (always) non-Gaussian with exponentially decaying tails. In fact, as illustrated in Figure~\ref{fig-ps-2}, the distributions for $\gamma \leq 0$ become highly skewed compared to the $\gamma >0$ case.

One can also use the simplified expressions of $Q(p,0)$ in Eq.~\eqref{mom-eq-6} to calculate the higher moments of work. For example, using Eq.~\eqref{Moments}, we obtain the second moment of $W$ to be
\begin{align}
\mom{W^2} \simeq 
\begin{cases}
\mom{W}^2+\frac{2k^2 v^2 L}{\gamma^3 D^2},~~~~~~~~~~~~~~\text{for }\gamma >0, \\
\frac{5v^4 L^4}{12 D^2},~~~~~~~~~~~~~~~~~~~~~~~~~~\text{for }\gamma =0, \\
2 \mom{W}^2-\frac{2kvL}{D \big| \gamma \big|} \mom{W} ,~~~~~~~\text{for  }\gamma <0.
\end{cases}
\label{mom-eq-8}
\end{align}

Once again, we see the emergence of different large-$L$ behaviours for different signatures of $\gamma$, stemming essentially from the same underlying physical reasoning as discussed in the context of the mean.
\section{Harmonic potential $(n=2)$}\label{harmonic}
We now consider the other solvable case of harmonic potential. For this case, the backward equation~\eqref{BackWard} takes the form
\begin{align}
    \centering
    \left[D\partial_{\xi_0}^2-{(k\xi_0+v)}\partial_{\xi_0}+kpv\xi_0) \right]Q(p,\xi_0)=0.
\end{align}
Solving this, we obtain
\begin{align}
    \centering
    & Q(p,\xi_0)  =\left[\mathbb{K}_1(p)~ H_{\Lambda(p)}\Big(\mathcal{R}(p,\xi_0)\Big)   +\mathbb{K}_2(p)~\,_1F_1\left(-\frac{\Lambda(p)}{2};\frac{1}{2};\mathcal{R}(p,\xi_0)^2\right)\right]e^{pv\xi _0},  \nonumber \\
 & \text{with   } \Lambda(p) = \frac{p v^2 ({D} p-1)}{k},~~~\text{and }
 \mathcal{R}(p,x) =\frac{k x-2 {D} p v+v}{\sqrt{2{D}k}}, \label{hamr-mom-gen}
\end{align}
\bluew{where $H_m(x)$ stands for the generalised Hermite polynomial with degree $m$ (where $m$ can be a real number)} and $\,_1F_1(m;1/2;x)$ is the Kummer confluent hypergeometric function. The other two functions $\mathbb{K}_1(p)$ and $\mathbb{K}_2(p)$ can be evaluated using the absorbing boundary conditions in Eq.~\eqref{Bcs} and we find
\footnotesize{\begin{align}
    \centering
    \mathbb{K}_1(p)&=\frac{e^{pvL} \, _1F_1\left(-\frac{\Lambda(p) }{2};\frac{1}{2};\mathcal{R}\left(p,L\right){}^2\right)-e^{-pvL} \, _1F_1\left(-\frac{\Lambda(p)
   }{2};\frac{1}{2};\mathcal{R}\left(p,-L\right){}^2\right)}{H_{\Lambda(p) }\Big(\mathcal{R}\left(p,-L\right)\Big) \, _1F_1\left(-\frac{\Lambda (p)
   }{2};\frac{1}{2};\mathcal{R}\left(p,L\right){}^2\right)-H_{\Lambda (p) }\Big(\mathcal{R}\left(p,L\right)\Big) \, _1F_1\left(-\frac{\Lambda (p)
   }{2};\frac{1}{2};\mathcal{R}\left(p,-L\right){}^2\right)},   \\
   \mathbb{K}_2(p)&=-\frac{e^{pvL} H_{\Lambda (p) }\Big(\mathcal{R}\left(p,L\right)\Big)-e^{-pvL} H_{\Lambda(p) }\Big(\mathcal{R}\left(p,-L\right)\Big)}{H_{\Lambda(p) }\Big(\mathcal{R}\left(p,-L\right)\Big) \, _1F_1\left(-\frac{\Lambda (p)
   }{2};\frac{1}{2};\mathcal{R}\left(p,L\right){}^2\right)-H_{\Lambda (p) }\Big(\mathcal{R}\left(p,L\right)\Big) \, _1F_1\left(-\frac{\Lambda (p)
   }{2};\frac{1}{2};\mathcal{R}\left(p,-L\right){}^2\right)}.
\end{align}}

\begin{figure}[t]
    \centering
    \includegraphics[scale=0.3]{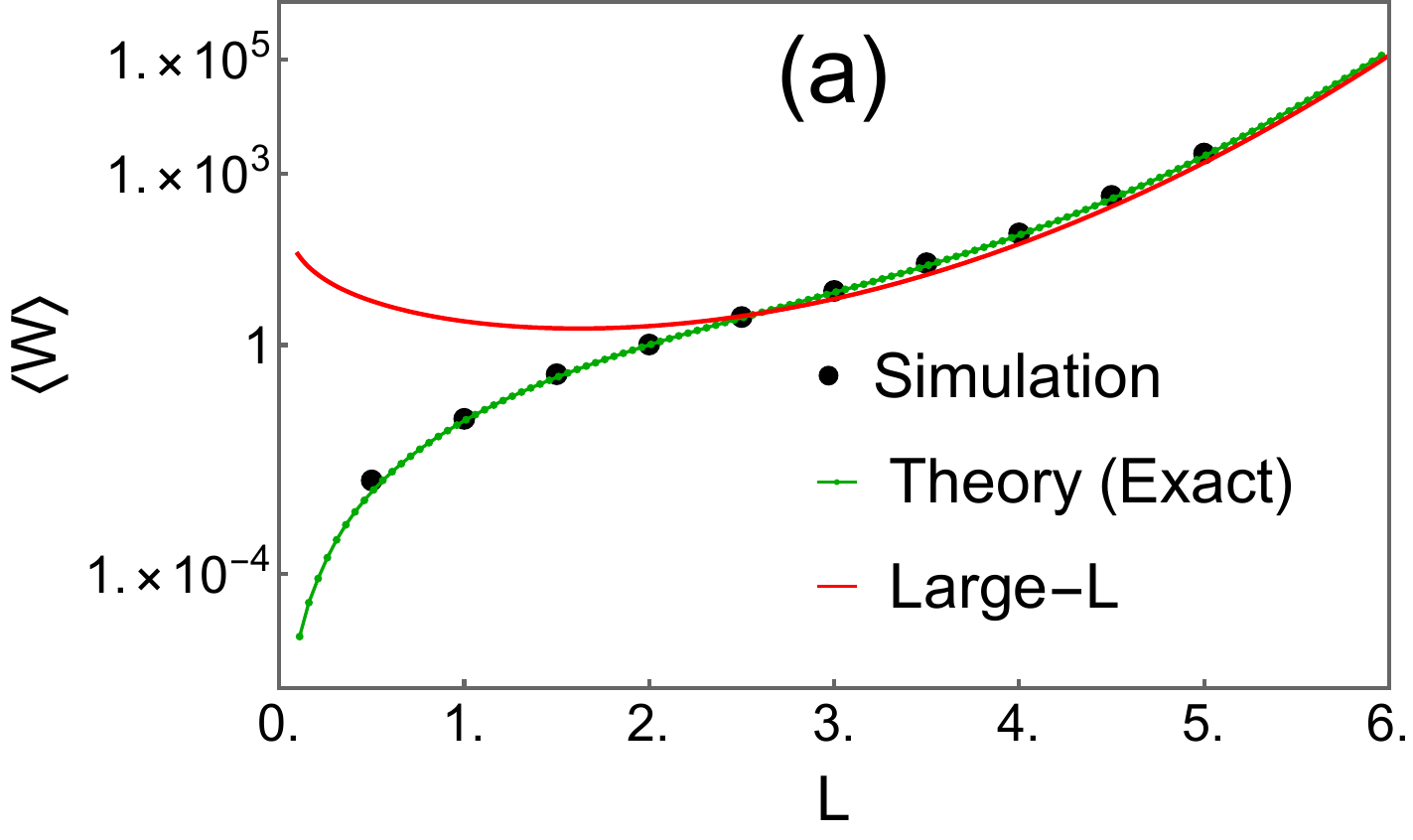}
    \includegraphics[scale=0.3]{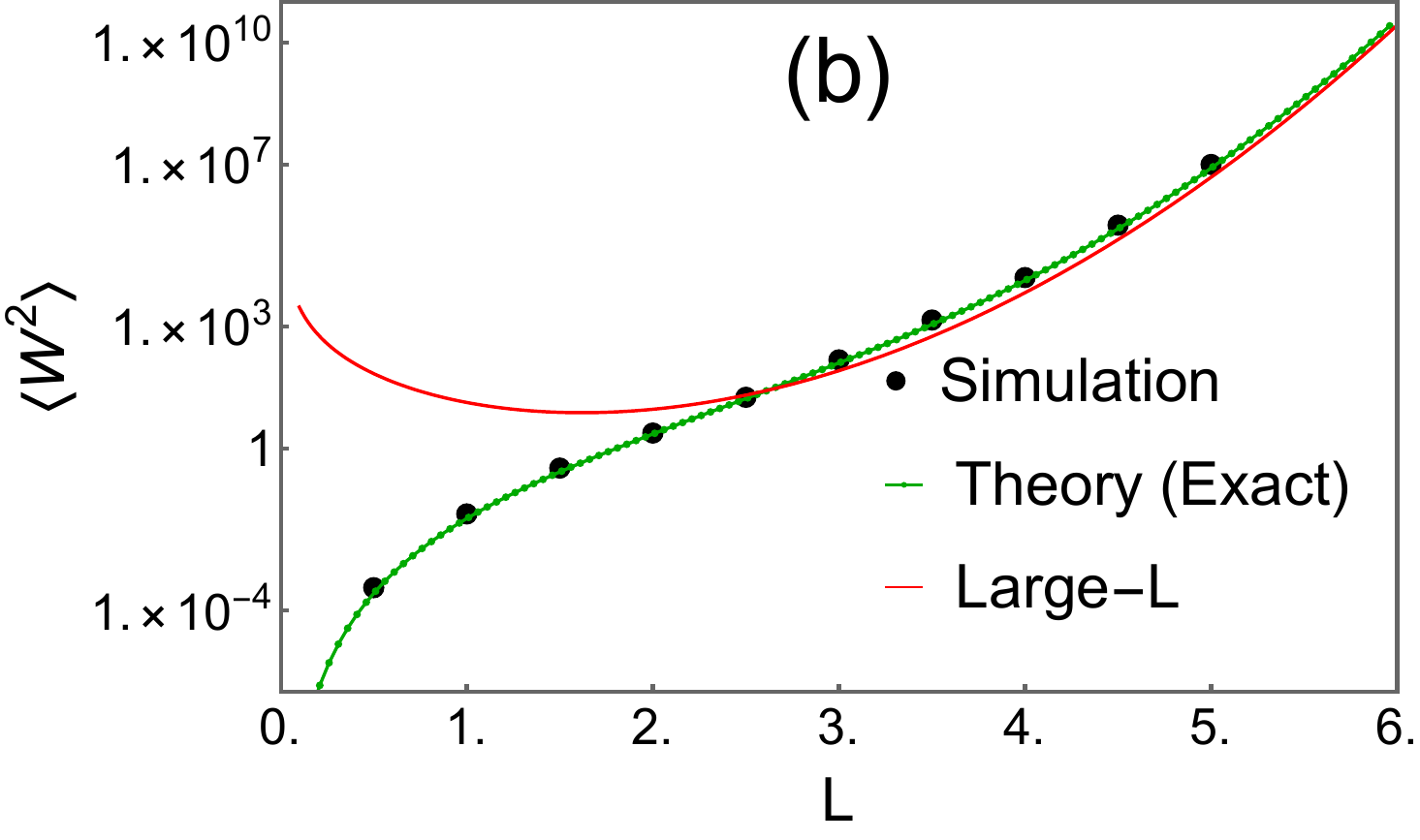}
    \caption{First two moments of the work done by the particle in presence of the harmonic potential $(n=2)$. In both panels, green curves are obtained by numerically differentiating the moment-generating function in Eq.~\eqref{hamr-mom-gen}, while red curves are the asymptotic results in Eqs.~\eqref{harm-eq-1} and \eqref{harm-eq-2}. We have chosen $v=k=D=1$.}
    \label{fig-ps-3}
\end{figure}

\normalsize{Now} by taking the derivative of $Q(p, \xi _0)$, we can find all moments of work done. In Figure~\ref{fig-ps-3}, we have illustrated this for the mean and the second moment of $W$ by numerically carrying out the derivative of Eq.~\eqref{hamr-mom-gen}. As seen from this plot, our results are consistent with the numerical simulations. Later, we will derive these moments exactly by using a slightly different (but related) method and show that for large $L$
\begin{align}
\mom{W} \sim L^{-1} \exp \left[ \frac{1}{D} \left( \frac{k L ^2}{2}-vL\right)\right], \label{harm-eq-1}
\end{align}
while the second moment of $W$ behaves as
\begin{align}
\mom{W^2} \simeq 2 \mom{W} ^2. \label{harm-eq-2}
\end{align}
\noindent We have verified these scaling behaviours in Figure~\ref{fig-ps-3} along with their exact counterparts. 

\begin{figure}[t]
    \centering
    \includegraphics[scale=0.45]{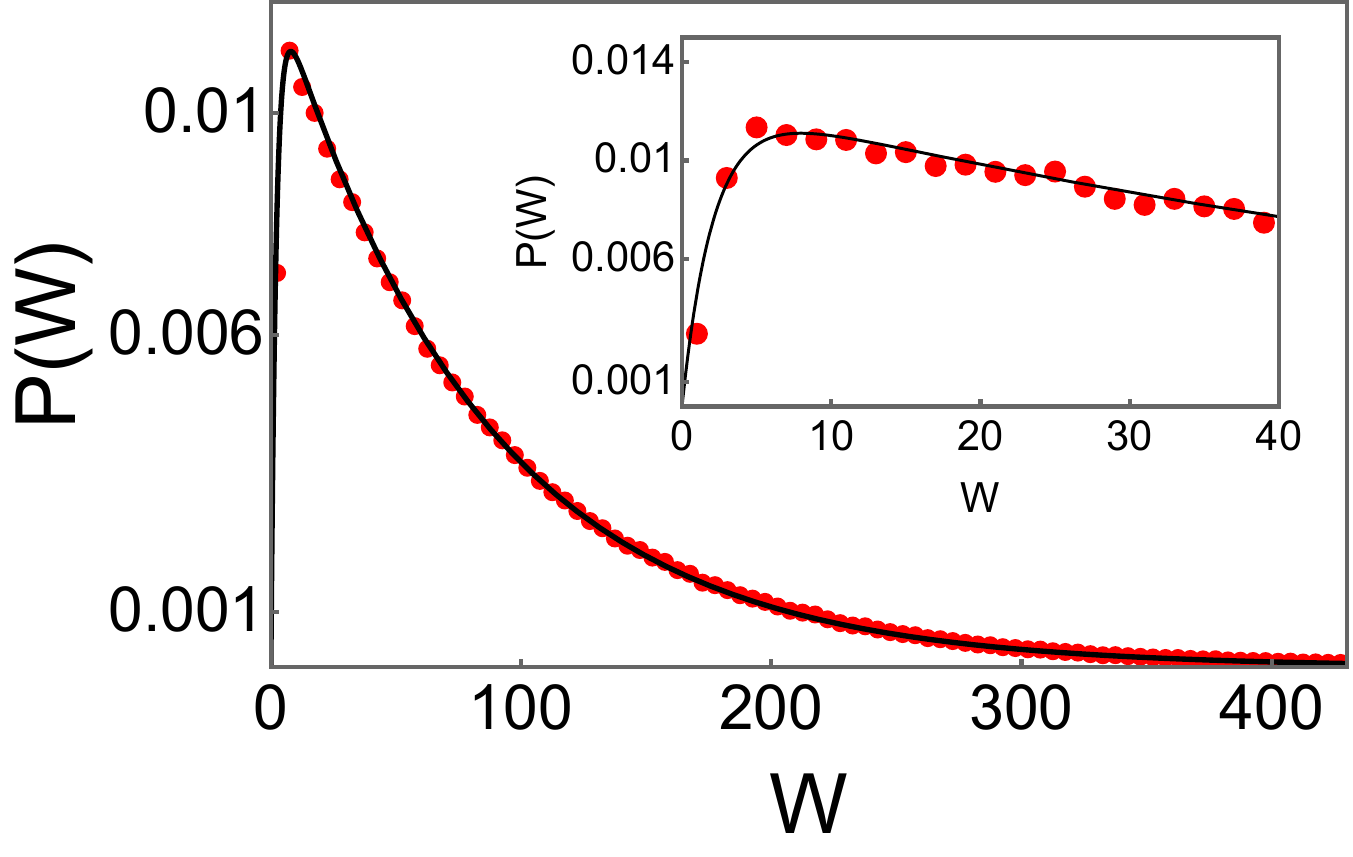}
    \caption{Probability distribution of the work done $W$ for the time-dependent harmonic potential $(n=2)$. Analytic expression in Eq.~\eqref{harm-eq-4} is plotted in black and the simulation data are shown in red. Inset shows the comparison for small values of $W$. Parameters chosen are $v=k=D=1$, $L=4$.}
    \label{fig-ps-4}
\end{figure}

Having looked at the first two moments, we next proceed to calculate the distribution of $W$. Obtaining the distribution using $Q(p,\xi_0)$ in Eq.~\eqref{hamr-mom-gen} analytically seems daunting. Nevertheless, in \ref{large-L-heuristic} we have provided a heuristic calculation that correctly gives the distribution for typical values of $W$ at large $L$. In particular, we find
\begin{align}
P(W) \simeq \frac{2 \exp \left(  -\frac{\mom{W}}{2 \mom{W}^2-\mom{W^2}}  W \right)       }{\sqrt{   2 \mom{W^2}-3 \mom{W}^2 }}~  \sinh \left( \frac{\sqrt{   2 \mom{W^2}-3 \mom{W}^2 }}{2 \mom{W}^2-\mom{W^2}} W \right), \label{harm-eq-4}
\end{align}
as $L$ becomes large. We have compared this expression with the numerical simulations in Figure~\ref{fig-ps-4} and we observe an excellent match between them. For large $W$, the distribution decays exponentially as $ \sim \exp \left( -W / \zeta _W \right)$ with the decay constant $\zeta _W = \frac{2 \mom{W}^2-\mom{W^2}}{\mom{W}-\sqrt{2 \mom{W^2} - 3 \mom{W}^2 }}$. This is in stark contrast with the case of the fixed observation time, where distribution of $W$ for $n=2$ case just turns out to be a Gaussian function \cite{PhysRevE.67.046102,ciliberto2017experiments}. In fact, the exponential decay turns out to be the hallmark property for the work distributions of all potentials with $n >1$ under a first-passage time as we discuss later.

\section{General $n$}\label{general}
Having examined the analytically tractable cases, we now shift our focus to deriving general results applicable to arbitrary $n~(>0)$. While solving the backward equation~\eqref{BackWard} directly for a general $n$ is a challenging task, we can utilize it to derive a differential equation governing the moments of $W$, which we can subsequently solve. These moments can then be employed to deduce the behaviour of the distribution of $W$ for large $L$ as done for the linear and harmonic cases previously. Writing the $m$-th moment of $W$ as $\mom{W^m(\xi _0)} = W_m (\xi _0)$, we take the derivative with respect to $p$ in Eq.~\eqref{BackWard} to obtain
\begin{align}
D \frac{d^2 W_m (\xi _0) }{d \xi _ 0 ^2}-\Big(k f_n(\xi _0)+v \Big) \frac{d W_m (\xi _0) }{d \xi _ 0 }= m v k f_n(\xi _0)  W_{m-1}(\xi _0), \label{gen-eq-1}
\end{align} 
where $W_0 (\xi _0) = 1$ and $f_n(\xi _0) =\sgn(\xi _0) \big|\xi _0 \big|^{n-1} $. For $m=1$, the solution of this equation is given by
\footnotesize{ \begin{align}
W_1 \left( \xi _0  \right) & =  \mathcal{B} + \mathcal{A} \sgn \left( \xi _0  \right) \int _{0}^{\big| \xi _0 \big|} dy~ \mathcal{G}_{ \sgn \left( \xi _0  \right)}(y)+\sgn \left( \xi _0  \right)  \frac{kv}{D}\int _{0}^{\big| \xi _0 \big|} dy~ \mathcal{G}_{ \sgn \left( \xi _0  \right)}(y) \int _{0}^{y}dz~\frac{z^{n-1}}{\mathcal{G}_{ \sgn \left( \xi _0  \right)}(z)},  \nonumber \\
&\text{with } \mathcal{G}_{\pm}(x) = \exp \left[ \frac{1}{D} \left(  \frac{k x^n}{n} \pm v x  \right) \right],
\label{gen-eq-2}
\end{align}}
\normalsize{and} the constants $\mathcal{A}$ and $\mathcal{B}$ can be computed using the condition $W_1 \left( \xi _0 = \pm L \right) = 0$. For $\xi _0 = 0$, the solution in Eq.~\eqref{gen-eq-2} reduces to
\footnotesize{\begin{align}
\mom{W} = \frac{kv}{D}\frac{\left(\int _{0}^{L} dy~ \mathcal{G}_{+}(y) \right)  \left(\int _{0}^{L} dy~ \mathcal{G}_{-}(y) \int _{0}^{y}dz~\frac{z^{n-1}}{\mathcal{G}_{-}(z)} \right)  -   \left(\int _{0}^{L} dy~ \mathcal{G}_{-}(y) \right)  \left(\int _{0}^{L} dy~ \mathcal{G}_{+}(y) \int _{0}^{y}dz~\frac{z^{n-1}}{\mathcal{G}_{+}(z)} \right)  }{ \int _{0}^{L} dy~ \left[ \mathcal{G}_{+}(y)+\mathcal{G}_{-}(y)     \right]   }. \label{gen-eq-3}
\end{align}}
\begin{figure}[t]
    \centering
    \includegraphics[scale=0.32]{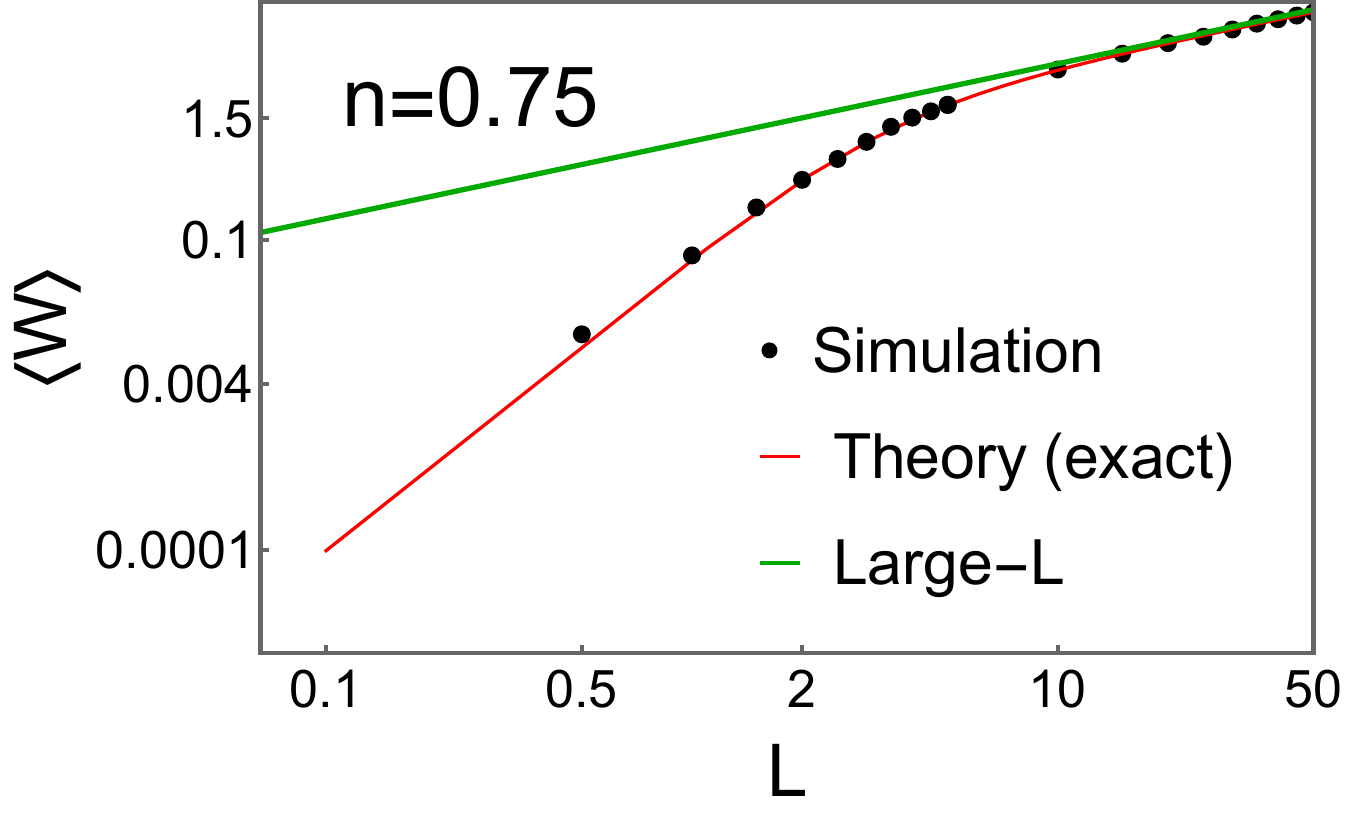}
    \includegraphics[scale=0.32]{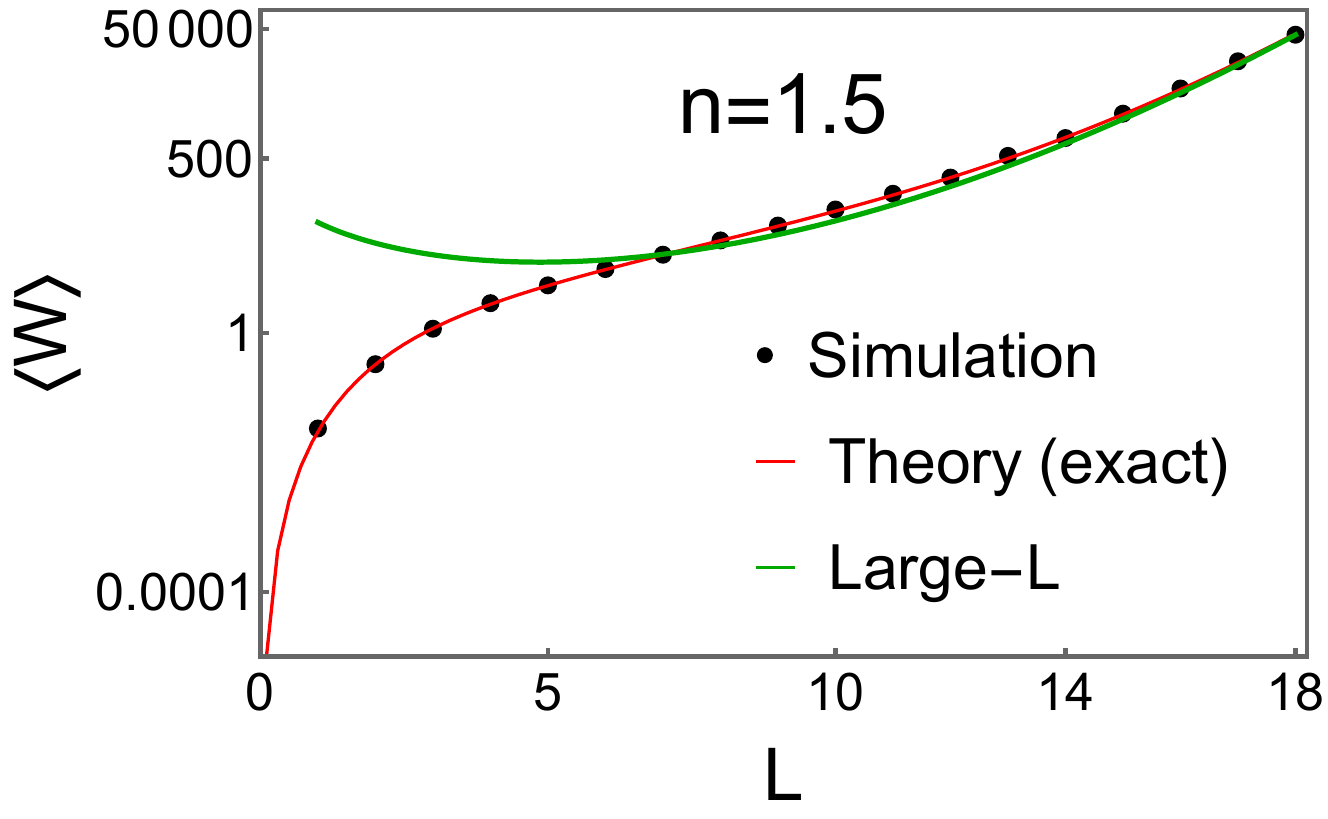}
    \caption{Illustration of the average work $\mom{W}$ performed by the particle in potentials with $n$ values $0.75$ and $1.5$. The theoretical plot is based on the expression in Eq.~\eqref{gen-eq-3}, presenting the exact calculation. In addition, the asymptotic large-$L$ results are given in  Eq.~\eqref{gen-eq-4}. Parameters chosen are $k=0.5,~v=D=1$.}
    \label{fig-ps-5}
\end{figure}

\noindent
\normalsize{This} is an exact expression of the mean valid for all $n$. Although analytically integrating in this expression  turns out to be challenging in general, one can simplify the expression and obtain the leading order behaviour of $\mom{W}$ in the large-$L$ limit. As shown in \ref{appen-mean-work}, this turns out to depend on the exponent $n$ and we get qualitatively different results depending on whether $n$ is greater or smaller than $1$. In particular, we have shown in \ref{appen-mean-work} that 
\begin{align}
\mom{W} \sim \begin{cases}
L^{1-n}~\exp \left[ \frac{1}{D} \left( \frac{k L^{n}}{n}-v L \right)  \right],~~~~~~\text{for }n>1, \\
L^{n},~~~~~~~~~~~~~~~~~~~~~~~~~~~~~~~~~~~~\text{for } n<1.
\end{cases} \label{gen-eq-4}
\end{align}
\bluew{For the marginal case $n=1$, these results are shown in Eq.~\eqref{mom-eq-3}. This can also be obtained from Eq.~\eqref{gen-eq-4} where $n \to 1^{-}$ gives the $\gamma >0$ case in Eq.~\eqref{mom-eq-3} while $n \to 1^{+}$ leads to the $\gamma <0$ case.} Next let us try to understand the difference between $n>1$ and $n<1$ cases physically. When the potential is sufficiently confining ($n>1$ case), the motion of the particle is heavily restricted due to the confining potential [see Figure \ref{sketch-fig}]. Thus, typically the particle will be located away from the absorbing walls at $\pm L$ and it will hit them very rarely. Therefore, for $n>1$, the first-passage time to reach the walls is typically high which in turn gives large value for the work. On the other hand, for $n<1$, while the potential is still confining, the corresponding force decays with increasing distance. Hence, in Eq.~\eqref{LangevinNew}, the drift $v$ term is always dominating beyond a certain distance. As a result, we get relatively smaller values of the first-passage time and hence the work.


In Figure~\ref{fig-ps-5}, we have compared the exact result in Eq.~\eqref{gen-eq-3} with the same obtained from the numerical simulations for $n<1$ (left panel) and $n>1$ (right panel). We observe an excellent agreement between the theory and the numerics for both cases. In addition to this, we have also presented a comparison of the large-$L$ expressions in  Eq.~\eqref{gen-eq-4} for the two cases. We find that Eq.~\eqref{gen-eq-4} converges to the exact result as we take larger values of $L$.


So far in this section, we have solved Eq.~\eqref{gen-eq-1} only for the mean $(m=1)$. But one can also utilize it to obtain the higher moments of $W$. A similar analysis for $m=2$ gives
\footnotesize{\begin{align}
\mom{W^2} = \frac{\left(\int _{0}^{L} dy~ \mathcal{G}_{+}(y) \right)  \left(\int _{0}^{L} dy~ \mathcal{G}_{-}(y) \int _{0}^{y}dz~\frac{z^{n-1} W_1(-z)}{\mathcal{G}_{-}(z)} \right)  -   \left(\int _{0}^{L} dy~ \mathcal{G}_{-}(y) \right)  \left(\int _{0}^{L} dy~ \mathcal{G}_{+}(y) \int _{0}^{y}dz~\frac{z^{n-1} W_1(z)}{\mathcal{G}_{+}(z)} \right)  }{  (2kv)^{-1} D \int _{0}^{L} dy~ \left[ \mathcal{G}_{+}(y)+\mathcal{G}_{-}(y)     \right]  }. \label{gen-eq-5}
\end{align}}
\normalsize{with} the asymptotic behaviour derived in \ref{appen-var} to be
\begin{align}
\mom{W^2} \simeq \begin{cases}
2\mom{W}^2,~~~~~~\text{for }n>1, \\
\mom{W}^2,~~~~~~~~\text{for } n<1.
\end{cases} \label{gen-eq-6}
\end{align}

\begin{figure}[t]
    \centering
    \includegraphics[scale=0.35]{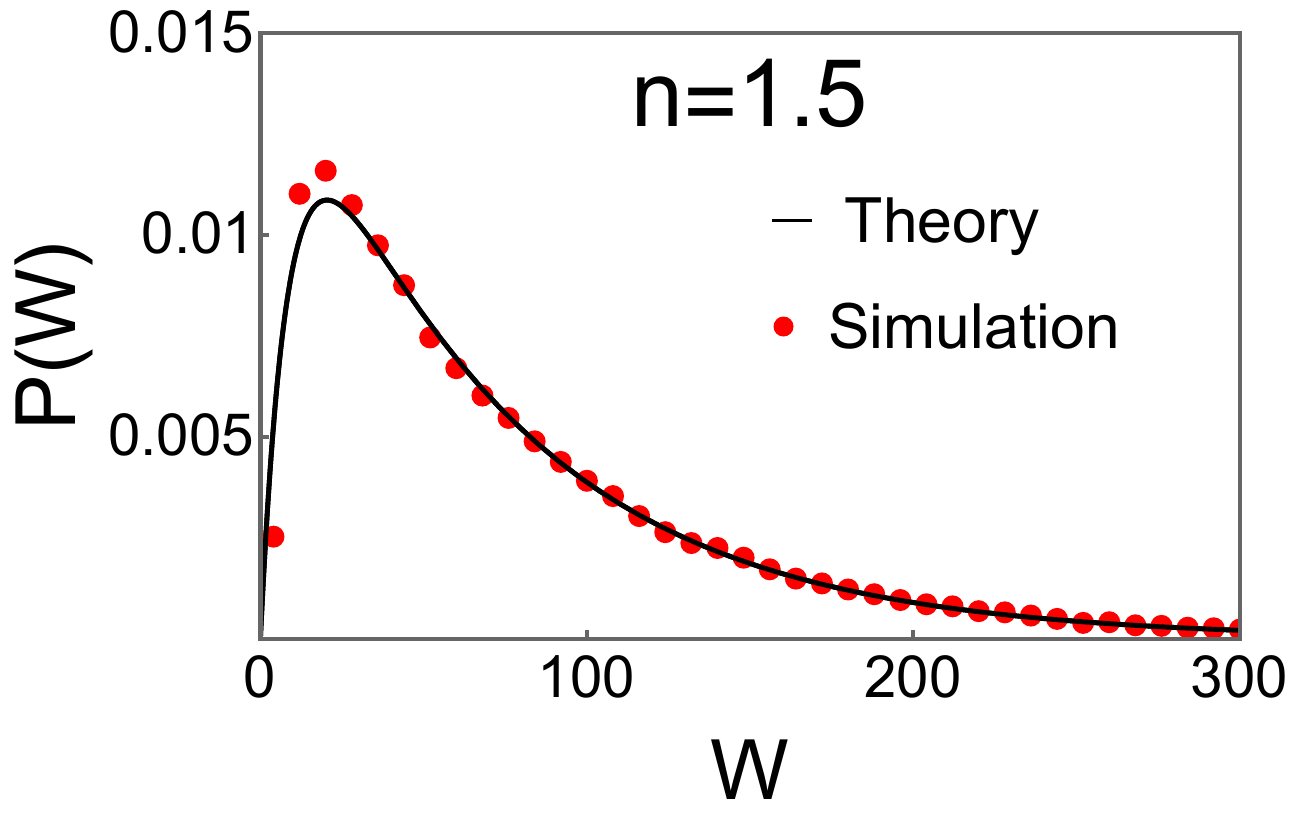}
    \includegraphics[scale=0.35]{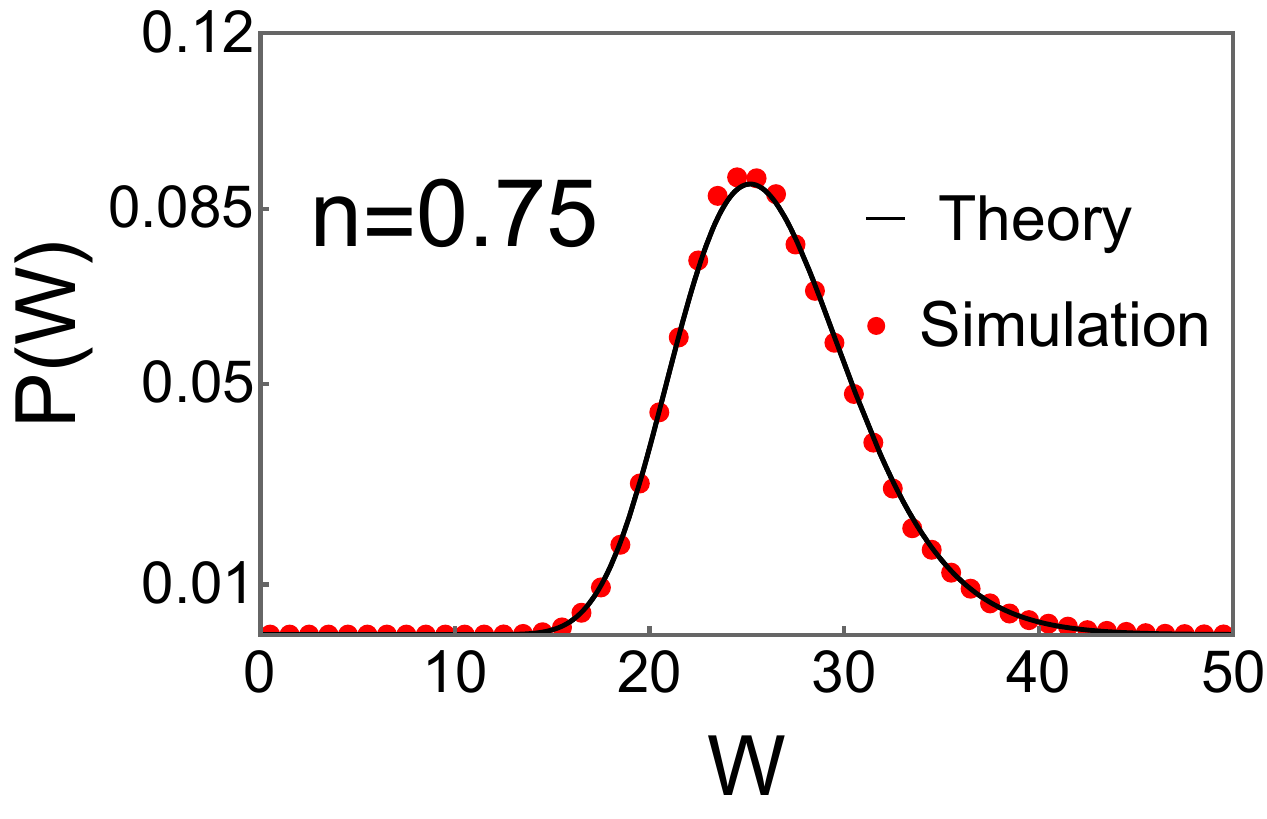}
    \caption{Comparison of the probability distribution of $W$ outlined in Eqs.~\eqref{gen-eq-8} and \eqref{gen-eq-11} is presented in the left panel for $n>1$ and $L=10$ and the right panel for $n<1$ and $L=100$. We have set the parameters as $k=0.5$, $v=D=1$ for this comparison using numerical simulations.}
    \label{fig-ps-6}
\end{figure}

\noindent
Note that these expressions only provide the leading order behaviour in $L$ and there are still sub-leading corrections that depend on model parameters $v$ and $k$. Interestingly, it turns out that one can use the first two moments to heuristically calculate the probability distribution describing the typical fluctuations of the work [as done before for $n=2$]. We refer to \ref{large-L-heuristic} for this calculation where we show
\begin{align}
P(W) \simeq \frac{2 \exp \left(  -\frac{\mom{W}}{2 \mom{W}^2-\mom{W^2}}  W \right)       }{\sqrt{   2 \mom{W^2}-3 \mom{W}^2 }}~  \sinh \left( \frac{\sqrt{   2 \mom{W^2}-3 \mom{W}^2 }}{2 \mom{W}^2-\mom{W^2}} W \right),~~~~~\text{for }n>1. \label{gen-eq-8}
\end{align}
We emphasize that this expression only describes the typical fluctuations of $W$ around the mean for $n>1$ and will not capture the regimes far away from the mean. In Figure~\ref{fig-ps-6} (left panel), we have compared this with the numerics for $n=1.5$. We notice an excellent agreement between them. Next, we address the case where $n<1$. As observed in the instance of $n=1$ with $\gamma >0$, one needs to solve the complete backward equation~\eqref{BackWard} in order to obtain $Q(p,0)$. While achieving this for $n=1$ is feasible, doing the same analytically for arbitrary $n<1$ turns out to be difficult. However, based on extensive numerical simulation, we find that for $n<1$ also, the distribution has the same form (around the mean) as for $n=1$ with $\gamma >0$ in Eq.~\eqref{mom-eq-5}. This leads us to assume
\begin{align}
P(W) \simeq \frac{1}{  \sqrt{  \mathbb{A}_1   W^3}}~\exp \left[  -\frac{\mathbb{A}_2}{W} \left(  W-\mathbb{A}_3 \right)^2 \right],~~~~\text{for }n<1, \label{gen-eq-9}
\end{align}
for large $L$. Here $\mathbb{A}_1$ and $\mathbb{A}_2$ and $\mathbb{A}_3$ are constants that depend on the parameters of the model. To compute them, we use the normalisation condition and the fact that the first two moments are given. We then obtain
\begin{align}
\mathbb{A}_1 &= \frac{2\pi \left( \mom{W^2}-\mom{W}^2 \right)  }{\mom{W}^3},~~~\mathbb{A}_2  = \frac{\mom{W}}{2 \left[ \mom{W^2}-\mom{W}^2  \right]},~~~~\mathbb{A}_3 = \mom{W}.\label{gen-eq-10}
\end{align}
Plugging these constants in Eq.~\eqref{gen-eq-9} gives
\begin{align}
P(W) \simeq \sqrt{\frac{  \mom{W}^3    }{  2\pi W^3\left( \mom{W^2}-\mom{W}^2 \right)   }}~\exp \left[-\frac{\mom{W} \left(  W-\mom{W} \right)^2  }{2W \left[ \mom{W^2}-\mom{W}^2  \right]}   \right], ~~~~\text{for }n<1. \label{gen-eq-11}
\end{align}
The right panel in Figure~\ref{fig-ps-6} shows a comparison of Eq.~\eqref{gen-eq-11} with the same obtained from the numerical simulations for $n=0.75$. An excellent match between them validates our results. In summary, we have derived the exact expressions for the first two moments and their large-$L$ forms in this section, applicable for all values of $n$. We then combined these expressions with a heuristic analysis to obtain the probability distribution of $W$ for large values of $L$ across all $n$ values. 

\bluew{Before ending this section, it is worth noting that although we have focused on the large-$L$ behaviour of the work done, one can also use the exact expressions given in Eqs.~\eqref{gen-eq-3} and \eqref{gen-eq-5} for computing the first two moments for small values of $L$. To calculate this, we first observe that for small $L$, one can perform the expansion $G_{\pm}(y) \simeq 1 + \frac{1}{D} \left( \frac{k y^n}{n} \pm vy\right)$ inside integrals in Eqs.~\eqref{gen-eq-3} and \eqref{gen-eq-5}. Now the integrations can be carried out analytically and we obtain
\begin{equation}
    \begin{rcases}
        \mom{W} &\simeq \frac{k v^2}{2 D^2}\frac{L^{n+2}}{(n+1)(n+2)},\\
        \mom{W^2} &\simeq ~\frac{2 k^2v^2 }{D^2}\frac{L^{2n+2}}{n(n+1)^2(n+2)},
    \end{rcases}
    ~~(\text{for small L).}
\end{equation}
Compared to the large-$L$ expressions, we find that both $\mom{W}$ and $\mom{W^2}$ exhibit algebraic scaling with $L$ for smaller values of $L$.}

\section{Relation with the first-passage time}\label{firstpassagetime}

So far, we have studied the statistics of the work done till the particle reaches one of the walls at $L_{\pm}(t)$ for the first time. Deploying a backward approach in terms of the co-moving variable $\xi(t)$, we have been able to compute the first two moments and the probability distribution of $W$ for large $L$. In the remaining part of our paper, we show that the fluctuations of the work done $W$ bear an interesting relation with the fluctuations of the first-passage time $t_f$ defined in Eq.~\eqref{FPT-new}. In order to see this, we first write a backward equation for the moments $T_{m}(\xi _0) = \mom{t_f^m(\xi_0)}$ of the first-passage time as done before in Eq.~\eqref{gen-eq-1} for the work done \cite{92dffd60-7698-3838-9cfd-f3a5e322ac6b, Majumdar-review-BF}
\begin{align}
D \frac{d^2 T_m (\xi _0) }{d \xi _ 0 ^2}-\Big(k f_n(\xi _0)+v \Big) \frac{d T_m (\xi _0) }{d \xi _ 0 }= -m  T_{m-1}(\xi _0), \label{new-FPT-eq-1}
\end{align} 
where $T_0(\xi _0) = 1$ and $f_n(\xi _0) =\sgn(\xi _0) \big|\xi _0 \big|^{n-1} $. Meanwhile the moments satisfy the boundary conditions $T_m \left( \xi _0 = \pm L \right) = 0$, since the particle gets instantly absorbed if its initial position coincides with one of the absorbing walls. Solving Eq.~\eqref{new-FPT-eq-1} in the same way as done in Section~\ref{general}, we obtain for $m=1$
\footnotesize{ \begin{align}
T_1 \left( \xi _0  \right) & =  \mathcal{B}_1 + \mathcal{B}_2 \sgn \left( \xi _0  \right) \int _{0}^{\big| \xi _0 \big|} dy~ \mathcal{G}_{ \sgn \left( \xi _0  \right)}(y)-  \frac{1}{D}\int _{0}^{\big| \xi _0 \big|} dy~ \mathcal{G}_{ \sgn \left( \xi _0  \right)}(y) \int _{0}^{y} \frac{dz}{\mathcal{G}_{ \sgn \left( \xi _0  \right)}(z)},
\label{new-FPT-eq-da1}
\end{align}}
\noindent
\normalsize{with} functions $\mathcal{G}_{\pm}(y)$ given in Eq.~\eqref{gen-eq-2}. Constants $\mathcal{B}_1$ and $\mathcal{B}_2$ can be evaluated using the boundary conditions $T_1 \left( \xi _0 = \pm L \right) = 0$. For $\xi _0 = 0$, the mean $ \mom{t_f}=T_1(0) $ is given by 
\footnotesize{\begin{align}
\mom{t_f} = \frac{1}{D}\frac{\left(\int _{0}^{L} dy~ \mathcal{G}_{+}(y) \right)  \left(\int _{0}^{L} dy~ \mathcal{G}_{-}(y) \int _{0}^{y}\frac{dz}{\mathcal{G}_{-}(z)} \right)  +  \left(\int _{0}^{L} dy~ \mathcal{G}_{-}(y) \right)  \left(\int _{0}^{L} dy~ \mathcal{G}_{+}(y) \int _{0}^{y}\frac{dz}{\mathcal{G}_{+}(z)} \right)  }{ \int _{0}^{L} dy~ \left[ \mathcal{G}_{+}(y)+\mathcal{G}_{-}(y)     \right]   }, \label{new-FPT-eq-2}
\end{align}}
\noindent

\normalsize{For} large $L$, the integrations in this expression can be approximately carried out as done for the work done. Proceeding similarly in \ref{appen-mean-work}, we find
\begin{align}
\mom{t_f} \sim \begin{cases}
L^{1-n}~\exp \left[ \frac{1}{D} \left( \frac{k L^{n}}{n}-v L \right)  \right],~~~~~~\text{for }n>1, \\
L,~~~~~~~~~~~~~~~~~~~~~~~~~~~~~~~~~~~~~\text{for } n<1.
\end{cases} \label{new-FPT-eq-3}
\end{align}
while for the marginal case of $n=1$, we obtain
\begin{align}
\mom{t_f} \simeq 
\begin{cases}
\frac{ L}{\gamma D},~~~~~~~~~~~~~~~\text{for }\gamma >0, \\
\frac{ L^2}{2 D},~~~~~~~~~~~~~~~\text{for }\gamma =0, \\
\frac{2k }{\gamma ^2 D^2 \mu}~e^{-\gamma L},~~~~\text{for  }\gamma <0,
\end{cases}~~~~ (n=1).
\label{new-FPT-eq-4}
\end{align}
Both Eqs.~\eqref{new-FPT-eq-3} and \eqref{new-FPT-eq-4} give the leading order behaviours in $L$. \bluew{Interestingly, for $n>1$, we see that the large-$L$ scaling of $\mom{t_f}$ is similar to $\mom{W}$ in Eq.~\eqref{gen-eq-4}. We believe this occurs because for large $t_f$, the rate of work done $W/t_f$ approaches a constant value (by the law of large numbers), and hence $W \sim t_f$. Therefore, the stochasticity of the total work is dominated by the stochasticity of the first-passage time.}

Next, one can solve Eq.~\eqref{new-FPT-eq-1} for for $m=2$ to show that the second moment $\mom{t_f^2} = \mom{T_m(\xi _0  = 0)}$ reads
\footnotesize{\begin{align}
\mom{t_f^2} = \frac{\left(\int _{0}^{L} dy~ \mathcal{G}_{+}(y) \right)  \left(\int _{0}^{L} dy~ \mathcal{G}_{-}(y) \int _{0}^{y}dz~\frac{ 	T_1(-z)}{\mathcal{G}_{-}(z)} \right)  + \left(\int _{0}^{L} dy~ \mathcal{G}_{-}(y) \right)  \left(\int _{0}^{L} dy~ \mathcal{G}_{+}(y) \int _{0}^{y}dz~\frac{ T_1(z)}{\mathcal{G}_{+}(z)} \right)  }{  2^{-1} D \int _{0}^{L} dy~ \left[ \mathcal{G}_{+}(y)+\mathcal{G}_{-}(y)     \right]  }. \label{new-FPT-eq-5}
\end{align}}

\normalsize{For} large $L$, one can again simplify this expression [see \ref{appen-var}] and obtain
\begin{align}
\mom{t_f^2} \simeq \begin{cases}
2\mom{t_f}^2,~~~~~~\text{for }n>1, \\
\mom{t_f}^2,~~~~~~~~\text{for } n<1,
\end{cases}
\label{nebahjd}
\end{align}
and for $n=1$, one gets 
\begin{align}
\mom{t_f^2} \simeq 
\begin{cases}
\mom{t_f}^2+\frac{2L}{D^2 \gamma ^3},~~~~~~~~~\text{for }\gamma >0, \\
\frac{5L^4}{12 D^2},~~~~~~~~~~~~~~~~~~~\text{for }\gamma =0, \\
2\mom{t_f}^2,~~~~~~~~~~~~~~~~\text{for  }\gamma <0.
\end{cases}~~~~ (n=1).
\end{align}
Once again in Eq.~\eqref{nebahjd}, we only obtain the leading order asymptotic expression in $L$. However, there will be sub-leading corrections that depend on parameters $v$ and $k$ and we do not delve into these corrections in Eq.~\eqref{nebahjd}.

\begin{figure}[t]
    \centering
    \includegraphics[scale=0.26]{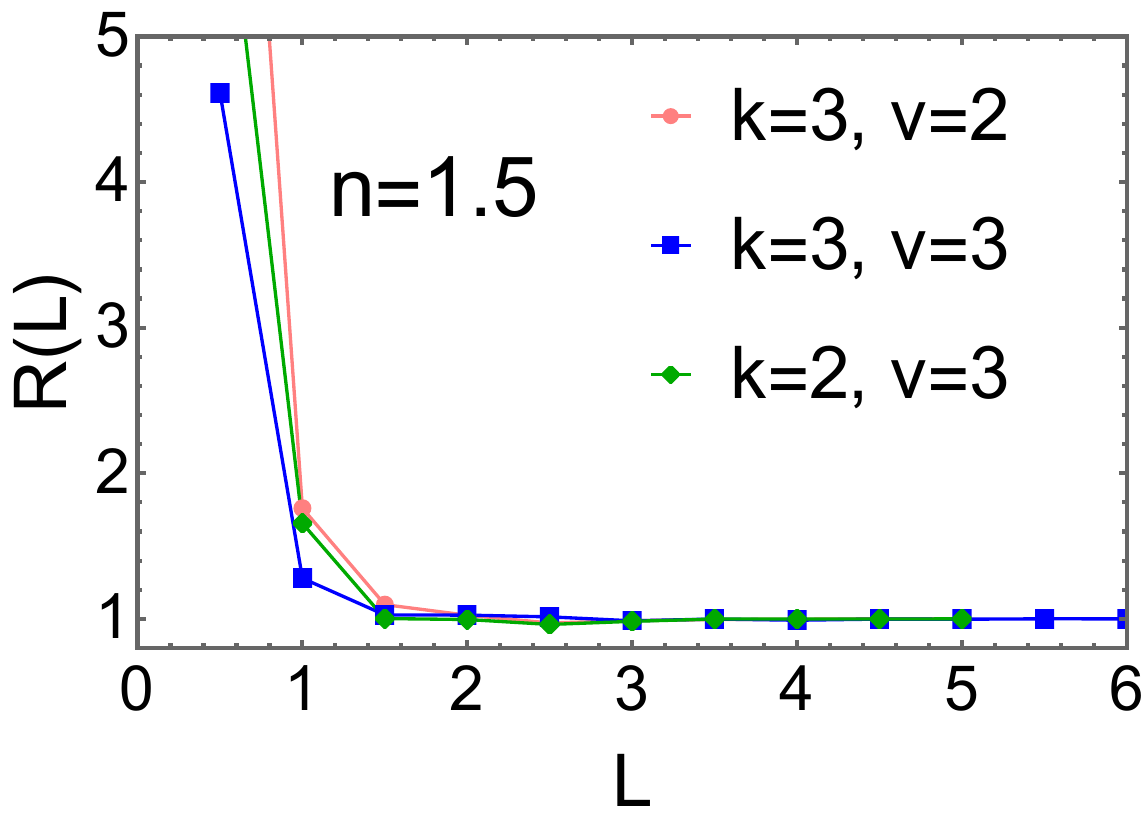}
    \includegraphics[scale=0.26]{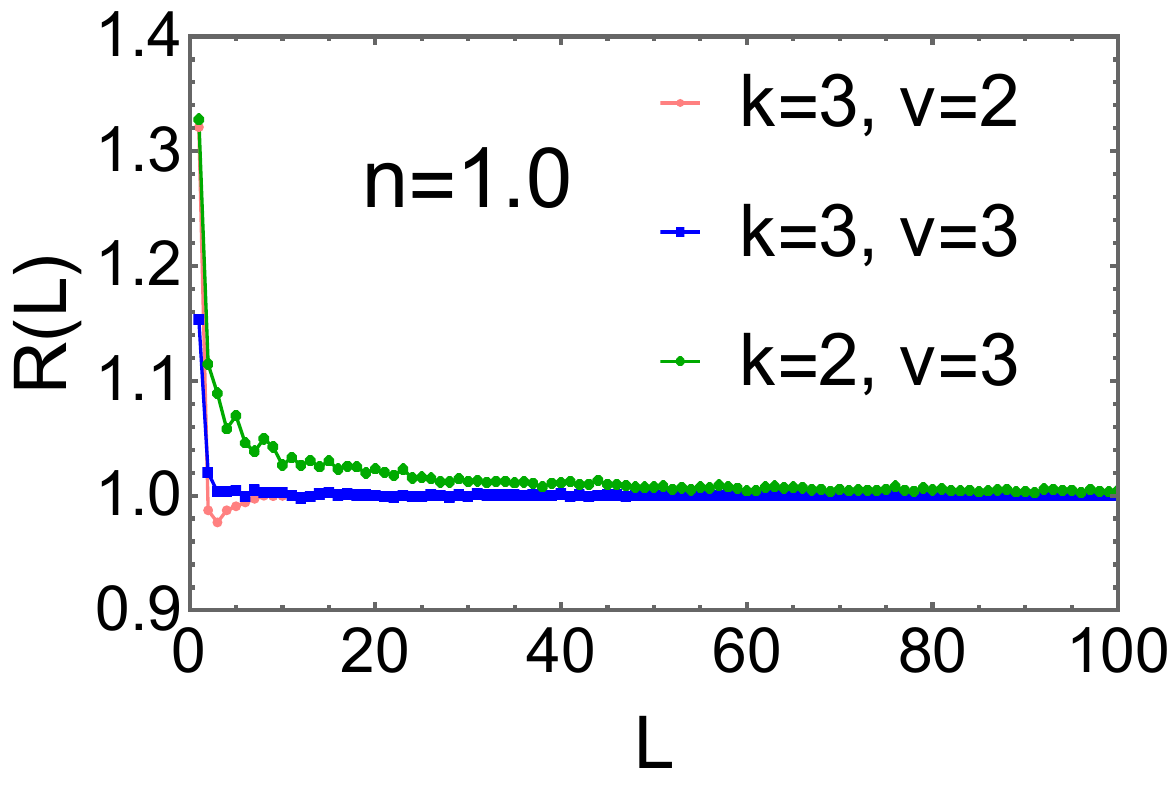}
    \includegraphics[scale=0.25]{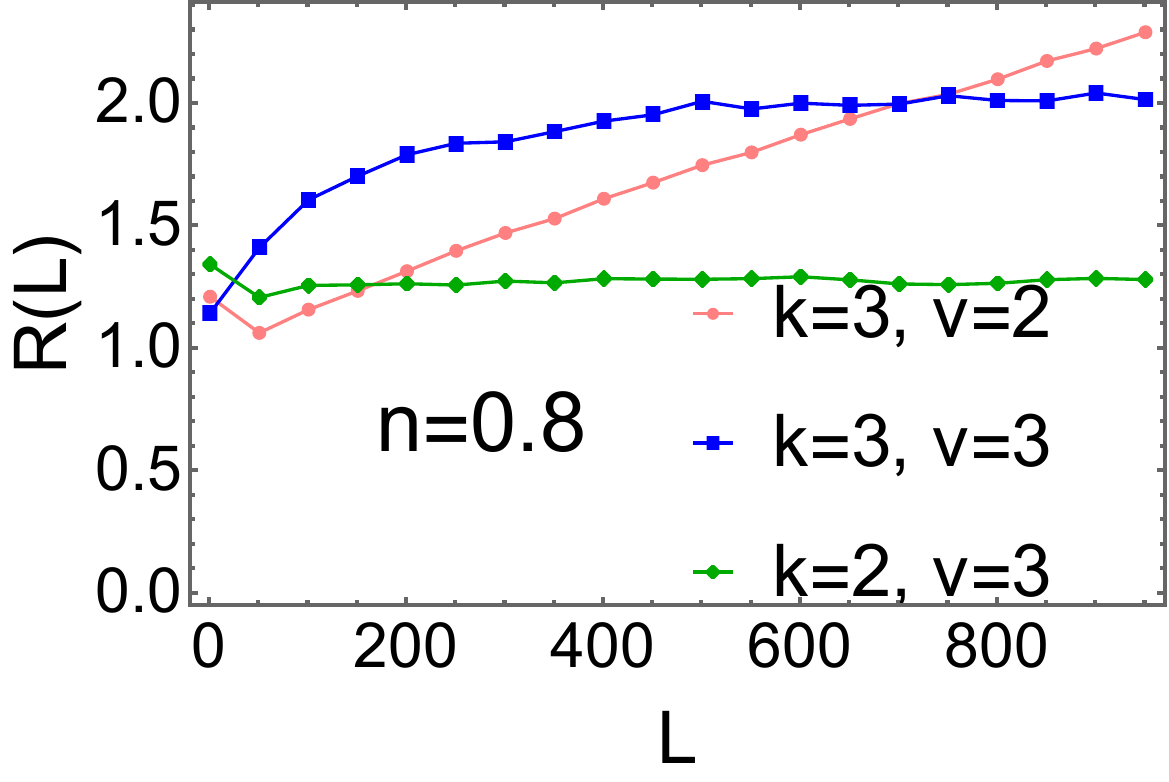}
    \caption{Simulation results for the ratio $R(L)$ defined in Eq.~\eqref{ratio-eqn} for three cases of $n>1$, $n=1$ and $n<1$. For every $n$, we have considered scenarios with $v<k$ (pink), $v=k$ (blue) and $v>k$ (green). }
    \label{ratio-fig}
\end{figure}

\bluew{Next, we set out to examine an interesting connection between the first-passage time and the work done. To this end, we define their respective coefficents of variation or the variabilities as}
\begin{align}
\text{CV}\big| _{W} = \sqrt{\frac{\mom{W^2}-\mom{W}^2}{\mom{W}^2}},~~~~\text{and }~~~\text{CV}\big| _{t_f} = \sqrt{\frac{\mom{t_f^2}-\mom{t_f}^2}{\mom{t_f}^2}},
\end{align}
and then study the following ratio 
\begin{align}
R(L) = \left( \frac{\text{CV}\big| _{W}}{\text{CV}\big| _{t_f}}\right) ^2. \label{ratio-eqn}
\end{align}
\bluew{As we show below, this ``variability ratio'' has intricate behaviors as a function of system size $L$ and the potential order $n$.}

For $n=1$, one can plug the analytic expressions of first and second moments of $W$ and $t_f$ derived in the previous sections and show that the ratio $R(L)$ converges to the value $1$ as the length $L$ becomes large for all values of $v$ and $k$. Similarly, for $n>1$,  one can use the asymptotic results $\mom{W^2} \simeq 2 \mom{W}^2$ and $\mom{t_f^2} \simeq 2 \mom{t_f}^2$ to show that both $\text{CV}\big| _{W} $ and $\text{CV}\big| _{t_f} $ individually go to $1$. This indicates that the ratio $R(L)$ also converges to $1$ for larger $L$ for $n>1$. However for $n<1$, an analytical calculation of $R(L)$ remains unlikely. As indicated in Eqs.~\eqref{gen-eq-6} and \eqref{nebahjd}, for the case of $n<1$, one needs to have the knowledge of the sub-leading corrections in $L$ within the expressions of $\langle W^2 \rangle$ and $\langle t_f^2 \rangle$ in order to compute the $CV$-s. Moreover, extracting these corrections analytically from the exact expression turns out to be difficult and our analysis in Eqs.~\eqref{gen-eq-6} and \eqref{nebahjd} cannot capture these details. 

In Figure \ref{ratio-fig}, we have plotted the ratio $R(L)$ as a function of the initial wall separation $L$ for different values of the exponent $n$. For $n \ge 1$, we find that $R(L)$ indeed saturates to the universal value equal to one at large $L$ independent of the model parameters $k$, $v$ and $D$. On the other hand, for $n<1$, the large-$L$ behaviour of $R(L)$ depends on these parameters. For instance, in Figure \ref{ratio-fig} [right panel], we see that the saturation values for $v=3,~ k=3$ (shown in blue) and $v=3,~k=2$ (shown in green) are completely different for $n=0.8$. On the other hand, for $v=2,~k=3$ (shown in pink), $R(L)$ increases monotonically at large $L$ and does not saturate to a constant value. This is possibly due to the fact that the saturation of $R(L)$ only takes place at large $L$, \emph{i.e.} for $L \gg L_{\text{th}}$, where $L_{\text{th}}$ is some threshold length. For a given $n$, the threshold $L_{\text{th}}$ also depends on the parameters $v$ and $k$. In Figure \ref{ratio-fig} [right panel], we find that $L_{\text{th}} \sim 50$ for the green curve $\left( k/v <1 \right)$ whereas $L_{\text{th}} \sim 500$ for the blue one $\left( k/v =1 \right)$. Thus, we see a consistent rise in the value of $L_{\text{th}}$ as the ratio $k/v$ starts to increase. Following this trend, we anticipate $L_{\text{th}}$ to be very large for the pink curve $\left( k/v >1 \right)$. Probing such large $L$ values in simulations is computationally expensive. Thus, we observe that $R(L)$ increases for the pink curve since we are still in the regime $L \ll L_{\text{th}}$. However, if we go to very large $L$ values (beyond what is considered in the figure), we anticipate $R(L)$ to converge to a constant value even for the pink curve $(v=2,~k=3)$.



\section{Conclusion}\label{conclusion}
In summary, we have analyzed the work statistics in a driven overdamped system in one dimension. In contrast to the conventional cases where the observables are usually measured upto a fixed time, here we measure the work upto a random first passage time that is conditioned on a certain criterion. We 
employ the Feynman-Kac path integral approach to compute the work functional in various set-ups consisting of potentials of different configurations. We provide a comprehensive analysis of the work fluctuations which shows a rich behaviour as a function of the potential strength, the external drive and the interval over which the particle moves. Furthermore, we showed an interesting symmetry relation between the signal-to-noise relation or the variability of the work and that of the first-passage time. To understand this phenomena better, we also delved deeper into the attributing physical conditions. Notably, our results illustrate a marked difference in the work statistics between systems driven up to a fixed time and a first-passage time. 
\bluew{Although we considered a fixed initial condition $x_0 = 0$ to calculate these results, our method can, in principle, be extended to a general distribution $P(x_0)$. In particular, we anticipate the large-$L$ expressions in Eqs.~\eqref{gen-eq-4} and \eqref{gen-eq-6} for work functionals and in Eqs.~\eqref{new-FPT-eq-3} and \eqref{nebahjd} for first-passage time to be valid even for a general $P(x_0)$ as long as the variance of $x_0$ does not scale with the initial separation $L$.} 

Moving forward, it would be interesting to study the ``\bluew{variability ratio}'' in other systems and to identify any \bluew{similar pattern as unveiled here}. There has been a myriad of studies in recent times in the field of stochastic thermodynamics involving various other thermodynamic observables such as injected power, dissipated heat or entropy production. It is only natural to study the same also within our set-up and will be pursued elsewhere. Furthermore, our model could also be useful to study existing thermodynamic bounds for first-passage-time problems \cite{gingrich2017fundamental,PhysRevLett.125.120604,PhysRevLett.124.040601,PhysRevX.7.011019,aghion2023thermodynamic}. \bluew{Another interesting direction is to investigate how our method can be extended to incorporate other potentials. For instance, we assumed that both the potential and the boundaries move at the same velocity $v$. It would be interesting to explore what happens if we relax this condition and allow them to move with different velocities. Concluding, we believe that our results can be tested using controlled optical trap experiments \cite{ciliberto2017experiments,proesmans2016brownian,nalupurackal2023towards} which, hopefully, will also unfold new research directions to this problem.}



\section*{Acknowledgement}
PS and KP acknowledge support from the European Union’s Horizon 2020 research and innovation program under the Marie Sklodowska-Curie grant agreement No. 847523 `INTERACTIONS’ and grant agreement No.~101064626 `TSBC' and from the Novo Nordisk Foundation (grant No. NNF18SA0035142 and NNF21OC0071284). IM and CF acknowledge the financial support from Brazilian agencies CNPq and FAPESP under grants 2021/03372-2, 2021/12551-8 and 2023/00096-0. AP acknowledges research support from the Department of Science and Technology, India, SERB Start-up Research Grant Number SRG/2022/000080 and Department of Atomic Energy, Government of India.

\appendix
\section{Large-$L$ behaviour of $Q(p,0)$ for the linear potential}
\label{appen-linear}
In this appendix, we provide the derivation of the simplified expression of the moment generating function $Q(p,0)$ for large $L$ as written in Eq.~\eqref{mom-eq-4}. For this, we first rewrite the exact expression of $Q(p,0)$ in Eq.~\eqref{mom-eq-1} as,
\begin{align}
Q(p,0) = \frac{ \mathcal{N}(p)  }{ \mathcal{D}(p)}, \label{appen-linear_eq_1}
\end{align}
where the functions $\mathcal{N}(p)$ and $\mathcal{D}(p)$ are given by
\small{\begin{align}
\mathcal{N}(p)& = \left[ \lambda _+(p) -\lambda _-(p)  \right] \left[ e^{-\sigma _+(p) L} -e^{-\sigma _-(p) L}  \right] + \left[ \sigma _+(p) -\sigma _-(p)  \right] \left[ e^{\lambda _-(p) L} -e^{\lambda _+(p) L}  \right], \\
\mathcal{D}(p) & = \left[  \lambda _-(p) - \sigma _+(p)  \right] e^{ 
\left[  \lambda _+(p) - \sigma _-(p)  \right] L }         +\left[  \sigma_+(p) - \lambda _+(p)  \right] e^{ 
\left[  \lambda _-(p) - \sigma _-(p)  \right] L }       \nonumber \\  
&    ~~~~~~~~~~~~~~+\left[  \lambda _+(p) - \sigma _-(p)  \right] e^{ 
\left[  \lambda _-(p) - \sigma _+(p)  \right] L }         +\left[  \sigma_-(p) - \lambda _-(p)  \right] e^{ 
\left[  \lambda _+(p) - \sigma _+(p)  \right] L } ,
\end{align}}
with $\lambda _{\pm}(p)$ and $\sigma _{\pm}(p)$ given in Eq~\eqref{lamb-eq}. In order to simplify this expression further, we first note that the work done by the particle typically attains a large positive value which scales either algebraically or exponentially with the length $L$, as seen from the expression of $\mom{W}$ in Eq.~\eqref{mom-eq-3}. In terms of $p$-variable, this translates to taking the $p \to 0$ limit in $Q(p,0)$. However, for small $p$, the leading order form of $Q(p,0)$ crucially depends on the signature of $\gamma$. Below we consider these different cases separately.
\subsection{$\gamma >0$ case}
For $p \to 0$ limit, we find that $\sigma _+ (p) \simeq \gamma $, $\lambda _ + (p) \simeq \mu $, whereas both $\sigma _- (p)$ and $\lambda _ -(p)$ are of order $\sim p $. Therefore, we find that $\mathcal{N}(p) \simeq - \gamma e^{\mu L} $ and $\mathcal{D}(p) \simeq -\gamma e^{\left[ \mu - \sigma _-(p) \right] L} $. Plugging this in Eq.~\eqref{appen-linear_eq_1}, we find
\begin{align}
Q(p,0) \simeq \exp \left[ \sigma _- (p)L \right],~~~~~\text{for }\gamma >0.
\end{align}
\subsection{$\gamma <0$ case}
On the other hand, for $\gamma <0$, we find $\sigma _- (p) \simeq \gamma $, $\lambda _ + (p) \simeq \mu $, while both $\sigma _+(p)$ and $\lambda _ -(p)$ are linear order in $ p $. This implies that to the leading order, we have 
\begin{align}
\mathcal{N}(p) &  \simeq \left[ \sigma _-(p)-\sigma _+(p)\right] e^{\lambda _+ (p) L}, \\
\mathcal{D}(p) &  \simeq  e^{\lambda _+ (p) L} \left[ e^{-\sigma _ +(p) L } \Big( \sigma _-(p) - \lambda _-(p)  \Big) +  e^{-\sigma _ -(p)L } \Big( \lambda _-(p) - \sigma _+(p)  \Big)   \right].
\end{align}
Inserting these expressions in Eq.~\eqref{appen-linear_eq_1} and performing the small $p$ expansion further, we obtain
\begin{align}
Q(p,0) \simeq \left[ 1+p \mom{W} \left( 1+ \frac{k v L}{D \big| \gamma \big|}p\right)  \right]^{-1},~~~~\text{for }\gamma <0.
\end{align}
\subsection{$\gamma = 0$ case}
For this case, we have $\lambda _ +(p) \simeq \mu $, $\lambda _-(p) \simeq v^2 p /\mu D $ and $\sigma _{\pm}(p) = \pm \sqrt{v^2 p/D}$. Plugging this in Eq.~\eqref{appen-linear_eq_1} and performing the small $\lambda _-$ expansion we find
\begin{align}
Q(p,0) \simeq \frac{\mu \left\{ \exp(2 \alpha L \sqrt{p}) -1  \right\} +2 \beta \sqrt{p} \exp(\alpha L \sqrt{p}) }{\mu \left\{ \exp(2 \alpha L \sqrt{p}) -1  \right\} + \beta \sqrt{p} ~ \left\{ \exp(2 \alpha L \sqrt{p}) +1  \right\} },~~~~~~\text{for }\gamma =0.
\end{align}

\section{Heuristic derivation of the distribution $P(W)$ for $n>1$}
\label{large-L-heuristic}
In this appendix, we will give a hand-waving derivation of the distribution of the work done in Eq.~\eqref{gen-eq-8} for $n>1$. To begin with, we first recall that for large $L$, the work done typically attains large positive values as indicated by Eq.~\eqref{gen-eq-4}. In terms of the moment generating function $Q(p,0)$, this corresponds to performing a small $p$ approximation. Indeed, for the case of linear potential, this led us to obtain a simplified expression of $Q(p,0)$ in Eq.~\eqref{mom-eq-4}. In fact, for all $n>1$, we find that the moment-generating function can be expressed as
\begin{align}
Q(p, 0) & \simeq 1- p \mom{W} + \frac{\mom{W^2}}{2} p^2, \\
& \simeq \left[ 1+ p \mom{W} + p^2 \left( \mom{W}^2 -\frac{\mom{W^2}}{2} \right)  \right] ^{-1}, \label{appen-gen-eq-7}
\end{align}
for large $L$ with moments $\mom{W} $ and $\mom{W^2} $ given in Eqs.~\eqref{gen-eq-3} and \eqref{gen-eq-5} respectively. Performing the inverse Laplace transformation, we then obtain
\begin{align}
P(W) \simeq \frac{2 \exp \left(  -\frac{\mom{W}}{2 \mom{W}^2-\mom{W^2}}  W \right)       }{\sqrt{   2 \mom{W^2}-3 \mom{W}^2 }}~  \sinh \left( \frac{\sqrt{   2 \mom{W^2}-3 \mom{W}^2 }}{2 \mom{W}^2-\mom{W^2}} W \right),~~~~~\text{for }n>1. \label{appen-gen-eq-8}
\end{align}
This expression has been presented in Eq.~\eqref{gen-eq-8} in the main text.


\section{Asymptotic behaviour of the mean work for large $L$}
\label{appen-mean-work}
Here, we derive the simplified expression of the mean work $\mom{W}$ for large $L$. The exact expression was derived in Eq.~\eqref{gen-eq-3} which we rewrite here as
\begin{align}
& \mom{W} = \frac{k v}{D}\frac{\mathcal{J}_+(L)~\mathcal{I}_-(L)-\mathcal{J}_-(L)~\mathcal{I}_+(L)}{\mathcal{J}_+(L)+\mathcal{J}_-(L)}, \label{apeen-mean-work-eq-1} \\
\text{with } & \mathcal{J}_{\pm}(L) = \int _{0}^{L}dy ~\mathcal{G}_{\pm}(y),~~\text{and}~~ \mathcal{I}_{\pm}(L) =\int _{0}^{L}dy ~\mathcal{G}_{\pm}(y) \int _{0}^{y}dz~\frac{z^{n-1}}{\mathcal{G}_{\pm}(z)}, \label{apeen-mean-work-eq-2}
\end{align}
where $\mathcal{G}_{\pm}(y)$ are given in Eq.~\eqref{gen-eq-2}. In order to find the large-$L$ behaviour of $\mom{W}$, we have to calculate the corresponding forms of the functions $\mathcal{J}_{\pm}(L)$ and $\mathcal{I}_{\pm}(L)$. These however turn out to depend on the exponent $n$ and we get different forms depending on whether $n$ is greater or smaller than $1$. Below we consider these two cases separately.
\subsection{$n>1$ case} 
\label{appen-mean-work-secngt1}
For $n>1$, both $\mathcal{G}_{\pm}(y)$ functions have faster than exponential growth for large argument. This indicates that both $\mathcal{J}_{\pm}(L)$ also grow with $L$. To see this, we write them as
\begin{align}
\mathcal{J}_{\pm}(L)& = \int _{0}^{L}dy~\exp \left[ \frac{1}{D} \left(  \frac{k y^n}{n} \pm v y  \right) \right], \\
& = L \int _{0}^{1}d \omega~\exp \left[ \frac{1}{D} \left(  \frac{k \omega^n L^n}{n} \pm v \omega L  \right) \right], ~~~~( \omega = y /L). \label{apeen-mean-work-eq-3}
\end{align}

For large $L$, the integrand has a maximum value at $\omega =1$. Moreover, this value increases with $L$. Therefore, for large $L$, the overall integration in Eq.~\eqref{apeen-mean-work-eq-3} will get major contribution from the vicinity of $\omega = 1$. Performing an expansion of the integrand in Eq.~\eqref{apeen-mean-work-eq-3} around $y = 1-u$ and keeping the leading order in $u$, we get the large-$L$ behaviour of $\mathcal{J}_{\pm}(L) $ as
\begin{align}
\mathcal{J}_{\pm}(L) & \sim L ~\exp \left[ \frac{1}{D} \left(  \frac{k  L^n}{n} \pm v  L  \right) \right] \times \int _{0}^{1} du~ \exp \left[ -\frac{u}{D} \left( k L^n \pm v L  \right)  \right], \\
& \sim L^{1-n} ~\exp \left[ \frac{1}{D} \left(  \frac{k  L^n}{n} \pm v  L  \right) \right].  \label{apeen-mean-work-eq-4}
\end{align}

We next turn to $\mathcal{I}_{\pm}(L) $. Rewriting their forms from Eq.~\eqref{apeen-mean-work-eq-2}
\begin{align}
\mathcal{I}_{\pm}(L) & = \int _{0}^{L}dy~~\exp \left[ \frac{1}{D} \left(  \frac{k y^n}{n} \pm v y  \right) \right] \int _{0}^y dz~z^{n-1}\exp \left[- \frac{1}{D} \left(  \frac{k z^n}{n} \pm v z  \right) \right] , \\
& =L\int _{0}^{1} d \omega~\exp \left[ \frac{1}{D} \left(  \frac{k \omega^n L^n}{n} \pm v \omega L  \right) \right]~\int _{0}^{\omega L} dz ~z^{n-1} \exp \left[- \frac{1}{D} \left(  \frac{k z^n}{n} \pm v z  \right) \right]. \label{apeen-mean-work-eq-5} 
\end{align}
Let us first analyse the integration over $z$ for large values of $L$. By changing the variable $u = k z^n /n D$, this integral can be rewritten as
\begin{align}
\int _{0}^{\omega L} dz ~z^{n-1} & \exp \left[- \frac{1}{D} \left(  \frac{k z^n}{n} \pm v z  \right) \right] \\
& \sim \int _{0}^{\frac{k w^nL^n}{n D}} du~ e^{-u} \sum _{i=0}^{\infty} \left( \mp \frac{v}{D} \right) ^i~\left( \frac{D n}{k} \right)^{\frac{i}{n}} \frac{u^{\frac{i}{n}}}{i!}, \\
& \sim  \sum _{i=0}^{\infty} \left( \mp \frac{v}{D} \right) ^i~\left( \frac{D n}{k} \right)^{\frac{i}{n}} \frac{\Gamma \left( 1+i/n\right)}{i!}, ~~~~~~~~\text{as } L \to \infty.
\end{align}
The sum converges for $n>1$. This implies that the integration over $z$ in Eq.~\eqref{apeen-mean-work-eq-5} becomes independent of $\omega L$ as $L$ becomes very large (for a given $\omega$) and $\mathcal{I}_{\pm}(L) $ accordingly takes the form
\begin{align}
\mathcal{I}_{\pm}(L) & \sim L\int _{0}^{1} d \omega~\exp \left[ \frac{1}{D} \left(  \frac{k \omega^n L^n}{n} \pm v \omega L  \right) \right], \\
& \sim  L^{1-n} ~\exp \left[ \frac{1}{D} \left(  \frac{k  L^n}{n} \pm v  L  \right) \right],
 \label{apeen-mean-work-eq-6} 
\end{align}
where the second line follows from Eq.~\eqref{apeen-mean-work-eq-4}. Combining this result with the expressions of $\mathcal{J}_{\pm}(L)$ in Eq.~\eqref{apeen-mean-work-eq-4} and plugging all of them in Eq.~\eqref{apeen-mean-work-eq-1}, we find
\begin{align}
\mom{W} \sim  L^{1-n} ~\exp \left[ \frac{1}{D} \left(  \frac{k  L^n}{n} - v  L  \right) \right],~~~~\text{for }n >1 \text{ and large }L.  \label{apeen-mean-work-eq-7} 
\end{align}

\subsection{$n <1$ case}
\label{sub-appen-nlt1}
Let us now derive the large-$L$ expression of the mean work $\mom{W}$ for $n <1$. Looking at Eq.~\eqref{apeen-mean-work-eq-1}, this again translates to calculating the functions $\mathcal{J}_{\pm}(L)$ and $\mathcal{I}_{\pm}(L)$. We first notice that even for $n<1$, $\mathcal{G}_+(y)$ has a faster than exponential growth for large $y$. Therefore, one can follow a similar approach as described in Section \ref{appen-mean-work-secngt1} for $\mathcal{J}_+(L)$ and $\mathcal{I}_+(L)$ and obtain
\begin{align}
\begin{cases}
& \mathcal{J}_{+}(L)  \sim \exp \left[ \frac{1}{D} \left(  \frac{k  L^n}{n} + v  L  \right) \right],  \\
& \mathcal{I}_{+}(L)  \sim \exp \left[ \frac{1}{D} \left(  \frac{k  L^n}{n} + v  L  \right) \right],
\end{cases} ~~~~~(\text{for large }L).
\label{apeen-mean-work-eq-9} 
\end{align}

On the other hand, $\mathcal{G}_-(y)$ for $n<1$ is a decaying function for large $y$ and requires a different approach. To this end, we recast $\mathcal{J}_-(L)$ as
\begin{align}
\mathcal{J}_-(L) = \int _{0}^{L}dy~e^{-v y/D}~\sum _{i=0}^{\infty} \left( \frac{k}{n D} \right)^{i} \frac{y^{ni}}{i!},
\end{align}
and carry out the integration over $y$ for $L \to \infty$ to yield
\begin{align}
\mathcal{J}_-(L) \sim \sum _{i=0}^{\infty} \left(  \frac{k}{D^{1-n} n ~v^n}  \right)^i ~\frac{\Gamma \left( 1+ ni \right)}{i!}. \label{apeen-mean-work-eq-10}
\end{align}
For $n<1$, this is a convergent sum. Hence $\mathcal{J}_-(L)$ becomes independent of $L$ for this case. Next we turn to $\mathcal{I}_-(L)$ whose explicit expression follows from Eq.~\eqref{apeen-mean-work-eq-2} as
\begin{align}
\mathcal{I}_-(L) = \int _{0}^{L}dy~~\exp \left[- \frac{1}{D} \left(  v y- \frac{k y^n}{n}   \right) \right] \int _{0}^y dz~z^{n-1}\exp \left[ \frac{1}{D} \left(  v z-\frac{k z^n}{n}  \right) \right].\label{apeen-mean-work-eq-11}
\end{align}
By changing the variables $y=\omega L $ and $z = u \omega L $, this expression becomes
\begin{align}
\mathcal{I}_-(L) = L^{n+1} \int _{0}^{1}d \omega ~\omega ^{n}\exp \left[ -\frac{\omega L}{D} \left( v-\frac{k}{n \left( \omega L\right)^{1-n}}\right) \right] \int _{0}^{1} du~u^{n-1}~\exp \left[ \frac{u \omega L}{D} \left( v-\frac{k}{n \left(u \omega L\right)^{1-n}}\right) \right] \nonumber.
\end{align}
For large $L$ and fixed $\omega$ and $u$, one can simplify this expression as
\begin{align}
\mathcal{I}_-(L) & \simeq L^{n+1} \int _{0}^{1}d \omega ~\omega ^{n}\exp \left[ -\frac{v \omega L}{D}  \right] \int _{0}^{1} du~u^{n-1}~\exp \left[ \frac{v u \omega L}{D}  \right] , \label{ghafvax01}\\
&\simeq L^{n+1} \int _{0}^{1}d \omega ~\omega ^{n}\exp \left( -\frac{v \omega L}{D}  \right)~\left( -\frac{D}{v \omega L}\right)^{n}\left[  \Gamma \left( n\right) -\Gamma \left(n,-\frac{v \omega L}{D} \right)  \right].
\end{align}
Using the approximate form of the gamma function as $\Gamma \left( n,-x \right) \stackrel{x \to \infty}{~\simeq~} (-x)^{n-1} ~e^x$ and then performing the integration over $\omega$ gives
\begin{align}
\mathcal{I}_-(L) \simeq \frac{D}{nv} L^{n},~~~\text{for large }L. \label{apeen-mean-work-eq-12}
\end{align}

Finally, using this result along with the results derived in Eqs.~\eqref{apeen-mean-work-eq-9} and \eqref{apeen-mean-work-eq-10} and inserting all of them in Eq.~\eqref{apeen-mean-work-eq-2}, we obtain the large-$L$ behaviour of the mean work as
\begin{align}
\mom{W} \simeq \frac{kv}{D}~ \mathcal{I}_-(L)\sim L^{n},~~~~\text{for }n <1.  \label{apeen-mean-work-eq-13}
\end{align}
Eqs.~\eqref{apeen-mean-work-eq-7} and \eqref{apeen-mean-work-eq-13} have been quoted in Eq.~\eqref{gen-eq-4} in the main text.

\section{Variance of the work done for large $L$}
\label{appen-var}
In this appendix, we will derive the simplified leading-order expression for the variance $\mom{W^2}$ of the work done for large $L$. The starting point is the exact expression in Eq.~\eqref{gen-eq-5} which we rewrite as
\begin{align}
\mom{W^2} = \frac{2 k v}{D}~\frac{\mathcal{J}_+(L)~\mathcal{Y} _-(L)-\mathcal{J}_-(L)~\mathcal{Y} _+(L)}{\mathcal{J}_+(L) + \mathcal{J}_-(L)}, \label{appen-var-eq-1}
\end{align}
where $\mathcal{J}_{\pm}(L)$ are given in Eq.~\eqref{apeen-mean-work-eq-2} and functions $\mathcal{Y}_{\pm}(L)$ are
\begin{align}
\mathcal{Y}_{\pm}(L) =\int _{0}^{L}dy ~\mathcal{G}_{\pm}(y) \int _{0}^{y}dz~W_1(\pm z)~\frac{z^{n-1}}{\mathcal{G}_{\pm}(z)}, \label{appen-var-eq-2}
\end{align}
with $\mathcal{G}_{\pm}(y)$ defined in Eq.~\eqref{gen-eq-2}. In order to find the large-$L$ expression of $\mom{W^2}$, we have to calculate the corresponding form of these functions. Meanwhile, we computed $\mathcal{J}_{\pm}(L)$ in \ref{appen-mean-work} and showed that they take different forms depending on whether the exponent $n$ is greater than or smaller than one. In the following, we carry out the same analysis for $\mathcal{Y}_{\pm}(L)$.
\subsection{$n>1$ case}
Let us write the explicit expression of $\mathcal{Y}_{\pm}(L)$
\begin{align}
\mathcal{Y}_{\pm}(L) & = \int _{0}^{L}dy~~\exp \left[ \frac{1}{D} \left(  \frac{k y^n}{n} \pm v y  \right) \right] \int _{0}^y dz~z^{n-1}~W_1(\pm z)~\exp \left[- \frac{1}{D} \left(  \frac{k z^n}{n} \pm v z  \right) \right] . \label{appen-var-eq-3}
\end{align}
Looking at this, we first notice that for $y$ integration, the integrand has a faster than exponential growth. This implies that at large $L$, this integration will be dominated predominantly by the larger values of $y$. On the other hand, the integration over $z$ is damped due to the presence of the decaying exponential term. Therefore, even though $y$ takes larger values, the integration over $z$ will still be dominated by small values of $z$ (more precisely $z \ll L$). This allows us to replace $W_1(\pm z) \simeq W_1(0)$ in Eq.~\eqref{appen-var-eq-3} and recast it as
\begin{align}
\mathcal{Y}_{\pm}(L) \simeq W_1(0) ~\mathcal{I}_{\pm}(L), \label{appen-var-eq-4}
\end{align}
where $\mathcal{I}_{\pm}(L)$ are given in Eq.~\eqref{apeen-mean-work-eq-2}. Inserting this in Eq.~\eqref{appen-var-eq-1} then yields
\begin{align}
\mom{W^2} \simeq  \frac{2 k v \mom{W}}{D}\frac{\mathcal{J}_+(L)~\mathcal{I}_-(L)-\mathcal{J}_-(L)~\mathcal{I}_+(L)}{\mathcal{J}_+(L)+\mathcal{J}_-(L)},
\end{align}
after which we use Eq.~\eqref{apeen-mean-work-eq-1} to obtain
\begin{align}
\mom{W^2} \simeq 2 \mom{W}^2,~~~~\text{for large }L.
\end{align}
\subsection{$n<1$ case}
We now turn to the other case of $n<1$. First recall that in \ref{sub-appen-nlt1}, we showed that the functions $\mathcal{J}_{\pm}(L)$ for this case scale as 
\begin{align}
\begin{cases}
& \mathcal{J}_{+}(L)  \sim \exp \left[ \frac{1}{D} \left(  \frac{k  L^n}{n} + v  L  \right) \right],  \\
& \mathcal{J}_{-}(L)  \sim \text{independent of }L,
\end{cases} ~~~~~(\text{for large }L).
\label{appen-var-eq-5} 
\end{align}
Similarly, for $\mathcal{Y}_+(L)$, we notice that $\mathcal{G}_{+}(y)$ is a growing function in $y$ even for $n<1$. This indicates that the approximations used in deriving Eq.~\eqref{appen-var-eq-4} for $\mathcal{Y}_+(L)$ will remain valid also for $n<1$ and we have
\begin{align}
\mathcal{Y}_+(L) \sim L^n~ \exp \left[ \frac{1}{D} \left(  \frac{k  L^n}{n} + v  L  \right) \right]. \label{appen-var-eq-55} 
\end{align}
However $\mathcal{Y}_-(L)$ has to be treated in a different way since $\mathcal{G}_-(y)$ is now a decaying function at large $y$. To this end, we first change the variables $y=\omega L $ and $z = u \omega L $ in Eq.~\eqref{appen-var-eq-2} and rewrite it as
\begin{align}
\mathcal{Y}_-(L) = & L^{n+1} \int _{0}^{1}d \omega ~\omega ^{n}\exp \left[ -\frac{\omega L}{D} \left( v-\frac{k}{n \left( \omega L\right)^{1-n}}\right) \right]  \nonumber \\
&  ~~~~~~~~\times      \int _{0}^{1} du~u^{n-1}~W_1(-u \omega L)
 \exp \left[ \frac{u \omega L}{D} \left( v-\frac{k}{n \left(u \omega L\right)^{1-n}}\right) \right].
\end{align}
Taking large $L$ limit for fixed $\omega $ and $u$, this simplifies to
\begin{align}
 \mathcal{Y}_-(L)  \simeq &  L^{n+1} \int _{0}^{1}d \omega ~\omega ^{n}\exp \left[ -\frac{v \omega L}{D}  \right] \int _{0}^{1} du~u^{n-1}~~W_1(-u \omega L)~\exp \left[ \frac{v u \omega L}{D}  \right]. \label{appen-var-eq-6} 
\end{align}
This expression is similar to Eq.~\eqref{ghafvax01} except for the presence of $W_1(-u \omega L)$ term. In \ref{subappen-work}, we show that
\begin{align}
W_1(-u \omega L) \simeq  \mom{W} (1-\omega^n u^n),~~~~\text{for large }L. \label{appen-var-eq-7} 
\end{align}
Plugging this expression in Eq.~\eqref{appen-var-eq-6} and performing the approximations as done in Eq.~\eqref{ghafvax01} yields
\begin{align}
\mathcal{Y}_-(L) \simeq \frac{\mom{W}}{2}~\mathcal{I}_-(L). \label{appen-var-eq-8} 
\end{align}
Recall that both $\mom{W}$ and $\mathcal{I}_-(L)$ scale as $\sim L^n$ for large $L$ [see Eq.~\eqref{apeen-mean-work-eq-13}] which implies that $\mathcal{Y}_-(L)  \sim L^{2n}$. 

We now have all functions needed to compute $\mom{W^2}$ in Eq.~\eqref{appen-var-eq-1}. Inserting the functions $\mathcal{J}_{\pm}(L)$ and $\mathcal{Y}_{\pm}(L)$ from Eqs.~\eqref{appen-var-eq-5}, \eqref{appen-var-eq-55} and \eqref{appen-var-eq-8} in Eq.~\eqref{appen-var-eq-1}, we finally obtain for $n<1$
\begin{align}
\mom{W^2} & \simeq \frac{k v \mom{W}}{D}~\mathcal{I}_-(L), \\
& \simeq \mom{W}^2,~~\text{[using Eq.~\eqref{apeen-mean-work-eq-13}]}. \label{appen-var-eq-9} 
\end{align}
\subsection{Proof of Eq.~\eqref{appen-var-eq-7}}
\label{subappen-work}
In the remaining part of this section, we provide a derivation of Eq.~\eqref{appen-var-eq-7} which was instrumental in deriving $\mom{W^2}$ for $n<1$ case. Following Eq.~\eqref{gen-eq-2}, we can write the mean work as
\begin{align}
W_1(-u \omega L) = \mom{W}-\mathcal{A}~\mathcal{J}_-(u  \omega L)-\frac{k v}{D}~\mathcal{I}_-(u  \omega L). \label{appen-var-eq-10} 
\end{align}
For $n<1$, $J_{-}(u \omega L)$ is independent of $L$ as indicated in Eq.~\eqref{appen-var-eq-5} whereas $I_{-}(u \omega L)$ scales as $\sim u^n \omega ^n L^n$ [shown in Eq.~\eqref{apeen-mean-work-eq-12}]. Using this in Eq.~\eqref{appen-var-eq-10}, we obtain
\begin{align}
W_1(-u \omega L) & \simeq \mom{W} - \frac{k}{n} u^n \omega^n L^n, \\
& \simeq \mom{W} \left( 1- u^n \omega ^n \right).
\end{align}
This proved the result in Eq.~\eqref{appen-var-eq-7}.

\section*{References}
\bibliographystyle{iopart-num}
\bibliography{Bib_new}

\end{document}